\pdfoutput=1
\documentclass[useAMS,usenatbib]{mn2e}
\usepackage{apjfonts}
\usepackage{graphicx}
\usepackage{amsmath}
\usepackage{amssymb}
\usepackage{ctable}
\usepackage{fixltx2e}
\usepackage{threeparttable}
\usepackage{hyperref}
\hypersetup{colorlinks=true,linkcolor=blue,citecolor=blue,filecolor=blue,urlcolor=blue}

\newcommand{\be}{\begin{equation}}
\newcommand{\ee}{\end{equation}}
\newcommand{\ba}{\begin{eqnarray}}
\newcommand{\ea}{\end{eqnarray}}

\newcommand{\Rvir}{R_{\rm vir}}
\newcommand{\Dvir}{\Delta_{\rm vir}}
\newcommand{\Mhalo}{{\rm M}_{\rm halo}}
\newcommand{\Rmax}{R_{\rm max}}
\newcommand{\Vmax}{V_{\rm max}}

\newcommand{\Ms}{{\rm M}_{\ast}}

\newcommand{\SFR}{{\rm SFR}}
\newcommand{\cm}{{\rm cm}}

\newcommand{\dex}{{\rm dex}}

\newcommand{\Msun}{{\rm M}_{\sun}}
\newcommand{\Zsun}{Z_{\sun}}

\newcommand{\mb}{m_b}
\newcommand{\mDM}{m_{\rm DM}}
\newcommand{\eb}{\epsilon_b}
\newcommand{\es}{\epsilon_{\rm star}}
\newcommand{\eDM}{\epsilon_{\rm DM}}

\newcommand{\nc}{n_{\rm th}}

\newcommand{\dd}{{\rm d}}
\newcommand{\LCDM}{$\Lambda$CDM}

\newcommand{\apj}{ApJ}
\newcommand{\apjl}{ApJ}
\newcommand{\apjs}{ApJS}
\newcommand{\mnras}{MNRAS}
\newcommand{\aap}{A\&A}

\newcommand{\nat}{Nature}
\newcommand{\rmxaa}{RMxAA}
\newcommand{\physrep}{Physics Reports}
\newcommand{\referee}[1]{{\color{black}#1}}

\title[High-$z$ galaxies on FIRE-2]
{Simulating galaxies in the reionization era with FIRE-2: galaxy scaling relations, stellar mass functions, and luminosity functions}

\author[X. Ma et al.]{
  \parbox[t]{1.0\textwidth}{
   Xiangcheng Ma,$^1$\thanks{E-mail: xchma@caltech.edu}
   Philip F. Hopkins,$^1$
   Shea Garrison-Kimmel,$^1$ \\
   Claude-Andr{\'e} Faucher-Gigu{\`e}re,$^2$
   Eliot Quataert,$^3$
   Michael Boylan-Kolchin,$^4$ \\
   Christopher C. Hayward,$^{5}$
   Robert Feldmann,$^6$ and
   Du{\v s}an Kere{\v s}$^7$ 
  }
  \vspace{5pt} \\
  $^1$TAPIR, MC 350-17, California Institute of Technology, Pasadena, CA 91125, USA \\ 
  $^2$Department of Physics and Astronomy and CIERA, Northwestern University, 2145 Sheridan Road, Evanston, IL 60208, USA \\
  $^3$Department of Astronomy and Theoretical Astrophysics Center, University of California Berkeley, Berkeley, CA 94720 \\
  $^4$The University of Texas at Austin, Department of Astronomy, 2515 Speedway, Stop C1400, Austin, Texas 78712-1205 \\
  $^5$Center for Computational Astrophysics, Flatiron Institute, 162 Fifth Avenue, New York, NY 10010, USA \\
  $^6$Institute for Computational Science, University of Zurich, Zurich CH-8057, Switzerland \\
  $^7$Department of Physics, Center for Astrophysics and Space Sciences, University of California at San Diego, 9500 Gilman Drive, La Jolla, CA 92093 \\
}

\pagerange{\pageref{firstpage}--\pageref{lastpage}}
\date{Draft version \today}

\begin{document}
\maketitle
\label{firstpage}

\begin{abstract}
We present a suite of cosmological zoom-in simulations at $z\geq5$ from the Feedback In Realistic Environments project, spanning a halo mass range $\Mhalo\sim10^8$--$10^{12}\,\Msun$ at $z=5$. We predict the stellar mass--halo mass relation, stellar mass function, and luminosity function in several bands from $z=5$--12. The median stellar mass--halo mass relation does not evolve strongly at $z=5$--12. The faint-end slope of the luminosity function steepens with increasing redshift, as inherited from the halo mass function at these redshifts. Below $z\sim6$, the stellar mass function and ultraviolet (UV) luminosity function slightly flatten below $\Ms\sim10^{4.5}\,\Msun$ (fainter than $\rm M_{1500}\sim-12$), owing to the fact that star formation in low-mass halos is suppressed by the ionizing background by the end of reionization. Such flattening does not appear at higher redshifts. We provide redshift-dependent fitting functions for the SFR--$\Mhalo$, SFR--$\Ms$, and broad-band magnitude--stellar mass relations. We derive the star formation rate density and stellar mass density at $z=5$--12 and show that the contribution from very faint galaxies becomes more important at $z>8$. Furthermore, we find that the decline in the $z\sim6$ UV luminosity function brighter than $\rm M_{1500}\sim-20$ is largely due to dust attenuation. Approximately 37\% (54\%) of the UV luminosity from galaxies brighter than $\rm M_{1500}=-13$ ($-17$) is obscured by dust at $z\sim6$. Our results broadly agree with current data and can be tested by future observations.
\end{abstract}

\begin{keywords}
galaxies: evolution -- galaxies: formation -- galaxies: high-redshift -- cosmology: theory 
\end{keywords}

\section{Introduction}
\label{sec:intro}
High-redshift galaxies are believed to be the dominant sources contributing to cosmic reionization (e.g. \citealt{cafg.2008:stellar.dominated.uvb,haardt.madau.2012:uvb,kuhlen.faucher.2012:reion,robertson.2013:reion.hudf12,robertson.2015:reion.planck}; however, see \citealt{madau.haardt.2015:reion.agn}). Current deep surveys using the {\it Hubble Space Telescope} have already put reliable constraints on the $z\geq5$ ultraviolet (UV) luminosity functions for galaxies brighter than $\rm M_{UV}=-17$ \citep[e.g.][]{mclure.2013:uvlf.z7to9.hudf12,schenker.2013:uvlf.z7to8.udf12,bouwens.2015:uvlf.z4to10,finkelstein.2015:uvlf.combined.field}, but the faint-end behavior of the UV luminosity function remains highly uncertain. These faint galaxies contribute a non-trivial fraction of the ionizing photons needed for reionization \citep[e.g.][]{finkelstein.2012:reion.candels,kuhlen.faucher.2012:reion,robertson.2013:reion.hudf12}, although their abundances are poorly understood.

Recently, \citet{livermore.2017:faint.galaxies} reported the detection of very faint galaxies of $\rm M_{UV}=-12.5$ at $z\sim6$ that are highly magnified by galaxy clusters in the Hubble Frontier Fields, after {performing} a novel analysis to remove the cluster light. They found a steep UV luminosity function down to $\rm M_{UV}=-13$ at $z\geq6$, implying sufficient numbers of faint galaxies to account for cosmic reionization. However, \citet{bouwens.2017:lensing.uncertainty,bouwens.2017:small.galaxy.sizes} later pointed out that the uncertain size distribution of high-redshift galaxies and the uncertain magnification model of the lensing clusters can have a large impact on the inferred faint-end luminosity functions in the Hubble Frontier Fields. The faint-end slope of the UV luminosity function fainter than $\rm M_{UV}=-15$ thus remains poorly constrained. 

Great efforts have also been made to measure the galaxy stellar mass functions at these redshifts \citep[e.g.][]{gonzalez.2011:smf.z4to7,duncan.2014:evolution.smf.z4to7,grazian.2015:stellar.mass.function,song.2016:stellar.mass.func.z4to8,stefanon.2017:hiz.optical.lf}. The stellar masses of high-redshift galaxies are usually derived from single-band photometry using empirical relations. Such relations are calibrated against spectral energy distribution (SED) fitting using limited rest-frame optical data for a small sample of galaxies at these redshifts. These relations tend to have large intrinsic scatter and suffer from systematic uncertainties of the underlying stellar population synthesis model. Therefore, the stellar mass functions reported by different authors have considerable discrepancies (e.g. figure 9 in \citealt{song.2016:stellar.mass.func.z4to8}). 

Consequently, the stellar mass--halo mass relation and the star formation efficiencies inferred from the stellar mass measurements at these redshifts are also very uncertain. For example, \citet{finkelstein.2015:stellar.baryon.frac} reported an increasing stellar mass to halo mass ratio with increasing redshift, whereas \citet{stefanon.2017:hiz.optical.lf} found no evolution of this ratio at these redshifts. Another related question is to understand the stellar mass growth histories of galaxies at these redshifts. This is not only useful for constraining the total ionizing photon emissivity at the epoch of reionization, but also essential for understanding galaxy populations at lower redshift -- both dwarf galaxy abundances in the Local Group \citep[e.g.][]{mbk.2015:lg.time.machine} and stellar mass functions in local galaxy clusters \citep[e.g.][]{lv.2014:sfh.model.in.dm.halo}.

The {\it James Webb Space Telescope} (JWST, scheduled for launch in 2020) and the next generation of ground-based telescopes will make it possible to study $z\geq5$ galaxies in more detail. Future observations of galaxies in the reionization era will provide substantial data for high-spatial-resolution deep imaging at the rest-frame optical bands, as well as spectroscopic measurements probing the physical conditions of the interstellar medium (ISM) in these galaxies. This may help resolve many current open questions in the field, such as the faint-end slope of the luminosity function, more robust determination of stellar mass, understanding the stellar populations in high-redshift galaxies and their contribution to cosmic reionization \citep{leitherer.2014:rotation.sb99,topping.shull.2015:rotation.ion,choi.2017:rotate.massive.star.pop,stanway.2017:massive.star.iau.proc}, etc. Therefore, it is necessary from a theoretical point of view to make more realistic predictions of galaxy properties at these redshifts. 

Currently there are two broad categories of cosmological simulations of galaxy formation at the epoch of reionization. High-resolution cosmological radiation-hydrodynamic simulations, with a detailed set of baryonic physics, including primordial chemistry and molecular networks, can simultaneously model the formation of first stars and galaxies and the local reionization history \citep[e.g.][]{wise.2014:reion.esc.frac,chen.2014:highz.scaling.relation,oshea.2015:renaissance.uvlf,paarde.2015:low.esc.frac}. Such calculations are usually computationally expensive and thus carried out in a small cosmological volume. They generally focus on the formation of Population III (Pop III) stars and low-mass galaxies (in halos below $\Mhalo\sim10^9\,\Msun$) at relatively high redshifts ($z\gtrsim10$). \referee{These types of simulations have been used to predict the scaling relations of high-redshift, low-mass galaxies \citep[e.g. the stellar mass--halo mass relation, gas fraction, mass--metallicity relation, etc.;][]{chen.2014:highz.scaling.relation}, ionizing photon escape fractions from these small galaxies and their importance for cosmic reionization \citep[e.g.][]{paarde.2015:low.esc.frac,xu.2016:renaissance.galaxies}, their spectral properties and detectability with {\em JWST} \citep[e.g.][]{barrow.2017:highz.galaxy.spectra}, and the faint-end ($\rm M_{UV}>-14$) UV luminosity functions at these redshifts \citep[e.g.][]{oshea.2015:renaissance.uvlf}.}

On the other hand, there are also large-volume cosmological simulations at relatively low resolution using empirically-calibrated models of star formation and stellar feedback {\citep[e.g.][]{feng.2016:bluetides.first.galaxy,gnedin.2016:croc.faint.end.uvlf,ocvirk.2016:coda.simulation,finlator.2017:minimum.halo.mass,pawlik.2017:aurora.reionization}}. \referee{Simulations of this nature broadly reproduce the observed galaxy populations, stellar mass functions, UV luminosity functions \citep[e.g.][]{gnedin.2016:croc.faint.end.uvlf,wilkins.2017:bluetides.simulation}, and the global reionization histories \citep[e.g.][]{ocvirk.2016:coda.simulation,pawlik.2017:aurora.reionization}. Forward modeling of galaxies in these simulations provide large samples of mock images and spectra that can be directly confronted with {\em JWST} \citep[e.g.][]{wilkins.2016:bluetides.photometry,zackrisson.2017:spectral.first.gal}. However, these simulations tend to have mass resolution $\gtrsim10^5\,\Msun$. Therefore, they are not able to capture the small-scale physics and the detailed structures in galaxies, which can be important for questions such as understanding the escape of ionizing photons \citep[e.g.][]{ma.2015:fire.escfrac}.} Also, some galaxy formation models calibrated to observations in the local universe struggle to reproduce observed galaxy properties at intermediate redshifts ($z\sim2$--3), such as star formation histories, metallicities, etc. \citep[e.g.][]{ma.2016:fire.mzr,dave.2016:mufasa.method}. This is also a known problem in semi-analytic models of galaxy formation \citep[e.g.][]{lu.2014:sam.compare.candel}.

In this work, we introduce a new suite of cosmological `zoom-in' simulations at $z\geq5$ in the $z=5$ halo mass range $\Mhalo\sim10^8$--$10^{12}\,\Msun$. We mainly focus on relatively massive (above the atomic cooling limit), Population II (Pop II) star-dominated galaxies in the redshift range $z=5$--12. Our simulations cover a range of galaxies that can be well probed by future observations using JWST and next-generation ground-based telescopes. The cosmological zoom-in technique allows us to simulate galaxies in a broad mass range without being limited to a fixed simulation volume. The resolution is adaptively chosen based on the mass of the system, but always much better than that of large-volume simulations. \referee{These are not the first cosmological zoom-in simulations at $z\geq5$: previous works using a similar technique have studied the escape fraction of ionizing photons \citep[e.g.][]{kimm.cen.2014:esc.frac}, galaxy properties and scaling relations \citep[e.g.][]{ceverino.2017:firstlight.project}, and the importance of stellar feedback for shaping these galaxies \citep[e.g.][]{yajima:2017:feedback.first.galaxy}.} Our work builds on these recent studies by increasing the resolution, expanding sample size, and most importantly including more detailed treatments for stellar feedback.

Our high-resolution cosmological zoom-in simulations use a full set of physically motivated models of the multi-phase ISM, star formation, and stellar feedback from the Feedback in Realistic Environments (FIRE) project\footnote{http://fire.northwestern.edu}. In a series of previous papers, these models have shown to successfully reproduce a variety of observed galaxy properties at lower redshifts \citep[e.g.][and references therein]{hopkins.2017:fire2.numerics}. Therefore, the new simulations presented in this paper are complementary to other state-of-the-art simulations in the field of galaxies in the reionization era.

This paper is the first in a series based on these new simulations, {focusing on galaxy properties, scaling relations, stellar mass functions, and luminosity functions at $z>5$. Our results complement previous predictions on the same topics using semi-analytic models \citep[e.g.][]{clay.2015:sam.high.redshift.uvlf,liu.2016:sam.uvlf.high.redshift,cowley.2018:sam.jwst.deep.survey} and cosmological simulations \citep[e.g.][]{jaacks.2012:steep.faint.end.uvlf,oshea.2015:renaissance.uvlf,yajima.2015:gal.property.z6to12,gnedin.2016:croc.faint.end.uvlf,ocvirk.2016:coda.simulation,xu.2016:renaissance.galaxies,wilkins.2017:bluetides.simulation}.} In Sections \ref{sec:ic} and \ref{sec:physics}, we describe the initial conditions and the physical ingredients used in the code. In Sections \ref{sec:definition} and \ref{sec:weights}, we construct the simulated catalog. In Section \ref{sec:galaxies}, we present the general properties of our simulated galaxies. In Sections \ref{sec:smf} and \ref{sec:lf}, we predict the stellar mass functions and luminosity functions from $z=5$--12. We discuss our results in Section \ref{sec:discussion} and conclude in Section \ref{sec:conclusion}.

We adopt a standard flat {\LCDM} cosmology with {\it Planck} 2015 cosmological parameters $H_0=68 {\rm\,km\,s^{-1}\,Mpc^{-1}}$, $\Omega_{\Lambda}=0.69$, $\Omega_{m}=1-\Omega_{\Lambda}=0.31$, $\Omega_b=0.048$, $\sigma_8=0.82$, and $n=0.97$ \citep{planck.2016:cosmo.param}. In this paper, we adopt a \citet{kroupa.2002:imf} initial mass function (IMF) from 0.1--$100\,\Msun$, with IMF slopes of $-1.30$ from 0.1--$0.5\,\Msun$ and $-2.35$ from 0.5--$100\,\Msun$. All magnitudes are in the AB system \citep{oke.gunn.1983:ab.system}.

\begin{table*}
\caption{Simulation details. Each simulation contains several galaxies in the zoom-in region. Properties below refer to the most massive (or the `target') galaxy.}
\begin{threeparttable}
\begin{tabular}{cccccccccc}
\hline
Name & $\Mhalo$ ($z=5$) & $\Ms$ ($z=5$) & $\Mhalo$ ($z=10$) & $\Ms$ ($z=10$) & $\mb$ & $\eb$ & $\es$ & $\mDM$ & $\eDM$ \\
 & [$\Msun$] & [$\Msun$] & [$\Msun$] & [$\Msun$] & [$\Msun$] & [pc] & [pc] & [$\Msun$] & [pc] \\
\hline
z5m09a & 2.4e+09 & 8.0e+05 & 2.3e+08 & 2.3e+04 & 119.3 & 0.14 & 0.7 & 6.5e+02 & 10 \\
z5m09b & 2.8e+09 & 5.9e+05 & 1.5e+08 & 5.2e+03 & 119.3 & 0.14 & 0.7 & 6.5e+02 & 10 \\
z5m10a & 6.7e+09 & 1.0e+07 & 1.4e+09 & 3.5e+05 & 954.4 & 0.28 & 1.4 & 5.2e+03 & 21 \\
z5m10b & 1.2e+10 & 1.6e+07 & 1.6e+09 & 1.1e+06 & 954.4 & 0.28 & 1.4 & 5.2e+03 & 21 \\
z5m10c & 1.3e+10 & 1.1e+07 & 1.0e+09 & 3.5e+05 & 954.4 & 0.28 & 1.4 & 5.2e+03 & 21 \\
z5m10d & 1.9e+10 & 2.0e+07 & 2.9e+08 & 2.4e+04 & 954.4 & 0.28 & 1.4 & 5.2e+03 & 21 \\
z5m10e & 2.4e+10 & 1.7e+07 & 6.8e+08 & 3.8e+05 & 954.4 & 0.28 & 1.4 & 5.2e+03 & 21 \\
z5m10f & 3.2e+10 & 1.1e+08 & 1.6e+09 & 3.4e+05 & 954.4 & 0.28 & 1.4 & 5.2e+03 & 21 \\
z5m11a & 4.1e+10 & 2.8e+07 & 1.4e+09 & 3.3e+05 & 954.4 & 0.28 & 1.4 & 5.2e+03 & 21 \\
z5m11b & 4.0e+10 & 9.2e+07 & 3.3e+09 & 3.4e+06 & 890.8 & 0.28 & 1.4 & 4.9e+03 & 21 \\
z5m11c & 7.5e+10 & 4.5e+08 & 8.5e+09 & 1.2e+07 & 7126.5 & 0.42 & 2.1 & 3.9e+04 & 42 \\
z5m11d & 1.4e+11 & 9.9e+08 & 2.5e+10 & 1.0e+08 & 7126.5 & 0.42 & 2.1 & 3.9e+04 & 42 \\
z5m11e & 2.4e+11 & 1.1e+09 & 1.3e+10 & 5.2e+07 & 7126.5 & 0.42 & 2.1 & 3.9e+04 & 42 \\
z5m12a & 4.4e+11 & 3.0e+09 & 2.3e+10 & 4.7e+07 & 7126.5 & 0.42 & 2.1 & 3.9e+04 & 42 \\
z5m12b & 8.5e+11 & 1.5e+10 & 3.2e+10 & 1.0e+08 & 7126.5 & 0.42 & 2.1 & 3.9e+04 & 42 \\
\hline
\end{tabular}
\begin{tablenotes}
\item Parameters describing the initial conditions for our simulations (units are physical):
\item (1) Name: Simulation designation.
\item (2) $\Mhalo$: Halo mass of the target halo at $z=5$ and its progenitor mass at $z=10$.
\item (3) $\Ms$: Stellar mass of the central galaxy in the target halo at $z=5$ and its progenitor mass $z=10$ (see Section \ref{sec:definition}).
\item (4) $\mb$: Initial baryonic (gas and star) particle mass in the high-resolution region. {A star particle loses about 25\% of its initial mass during its entire life due to mass return via supernovae and stellar winds.}
\item (5) $\eb$: Minimum Plummer-equivalent force softening for gas particles. Force softening for gas particles is adaptive. The gas inter-particle separation defined in \citet{hopkins.2017:fire2.numerics} is about $1.4\,\eb$.
\item (6) $\es$: Plummer-equivalent force softening for star particles (fixed in comoving units until $z=9$ and in physical units thereafter).
\item (7) $\mDM$: Dark matter particle mass in the high-resolution region.
\item (8) $\eDM$: Plummer-equivalent force softening for high-resolution dark matter particles (fixed in comoving units until $z=9$ and in physical units thereafter).
\end{tablenotes}
\end{threeparttable}
\label{tbl:sim}
\end{table*}

\section{The Simulations}
\label{sec:sim}
The simulations presented in this paper form a subsample of the FIRE project \citep[version 2.0, which we refer as FIRE-2;][]{hopkins.2017:fire2.numerics}. FIRE-2 is an updated version of the feedback implementations studied in a number of previous papers, which we refer as FIRE-1 \citep{hopkins.2014:fire.galaxy}. 

All FIRE-2 simulations are run using an identical version of the {\sc gizmo} code \citep{hopkins.2015:gizmo.code}\footnote{http://www.tapir.caltech.edu/{\textasciitilde}phopkins/Site/GIZMO.html}. We use the meshless finite-mass (MFM) method in {\sc gizmo} to solve the hydrodynamic equations. We refer to \citet{hopkins.2017:fire2.numerics} for details of the numerical methods and convergence tests of the FIRE-2 simulations, as well as their differences from FIRE-1 simulations. Other FIRE-2 simulations have already been presented and studied in recent papers \citep[e.g.][]{wetzel.2016:high.res.mw.letter,fitts.2017:field.dwarf.galaxy,sgk.2017:disk.model.satellite,elbadry.2018:gas.kinematics}. We describe the initial conditions of our simulated sample in Section \ref{sec:ic} and review the baryonic physics adopted in FIRE-2 briefly in Section \ref{sec:physics}.

\subsection{Initial conditions}
\label{sec:ic}
We run a set of dark matter-only cosmological boxes at low resolution to $z=5$, select target halos from the $z=5$ snapshots, and re-simulate these halos and the regions around them at much higher resolution with baryons using the well developed multi-scale cosmological `zoom-in' techniques \citep{katz.white.1993:xray.cluster,onorbe.2014:how.to.zoom}. The initial conditions of the parent boxes and the zoom-in simulations are generated at $z=99$ using the {\sc music} code \citep{hahn.abel.2011:music.code}, with {\em Planck} 2015 cosmological parameters.

We use three dark matter-only cosmological boxes of side-length 11, 22, and 43 comoving Mpc, respectively. We use the spherical overdensity-based Amiga Halo Finder \citep[{\sc ahf};][]{knollmann.knebe.2009:ahf.code} to identify halos in the $z=5$ snapshots, applying the redshift-dependent virial parameter from \citet{bryan.norman.1998:xray.cluster}, which leads to a virial overdensity $\Dvir\approx177$ (relative to background) for the redshift range we consider in this paper. We also checked the results against the six-dimensional phase-space halo finder {\sc rockstar} \citep{behroozi.2013:rockstar.code.paper} and found good agreements in halo mass functions. We randomly select target halos in the $z=5$ halo mass range $\Mhalo=2\times10^9$--$10^{12}\,\Msun$, requiring that there is no more massive halo within $5\Rvir$ from the target halo. \referee{This selection excludes 1/3 of the halos in the box\footnote{We also include other well-resolved halos in the zoom-in regions in our analysis (see Section \ref{sec:definition}). \referee{These halos live in the vicinity of a more massive halo (the target halo in the zoom-in region) by design. This will partially compensate the selection bias due to the isolation criteria above.}}.}
 
We identify zoom-in regions based on particles within $\sim3$--$5\Rvir$ of the targeted halo, and iterate to ensure zero mass contamination from low-resolution particles within $2\Rvir$ and less than 1\% contamination within $3\Rvir$ at $z=5$. There may be more than one halo in the zoom-in region, but the target halo is the most massive one by design. In Table \ref{tbl:sim}, we list all of our target halos studied in this paper, along with the halo mass and stellar mass of the central galaxy (see Section \ref{sec:definition} for details) at $z=5$ and $z=10$, and initial particle masses and minimum Plummer-equivalent force softening lengths of baryonic and high-resolution dark matter particles.

\begin{figure}
\centering
\includegraphics[width=\linewidth]{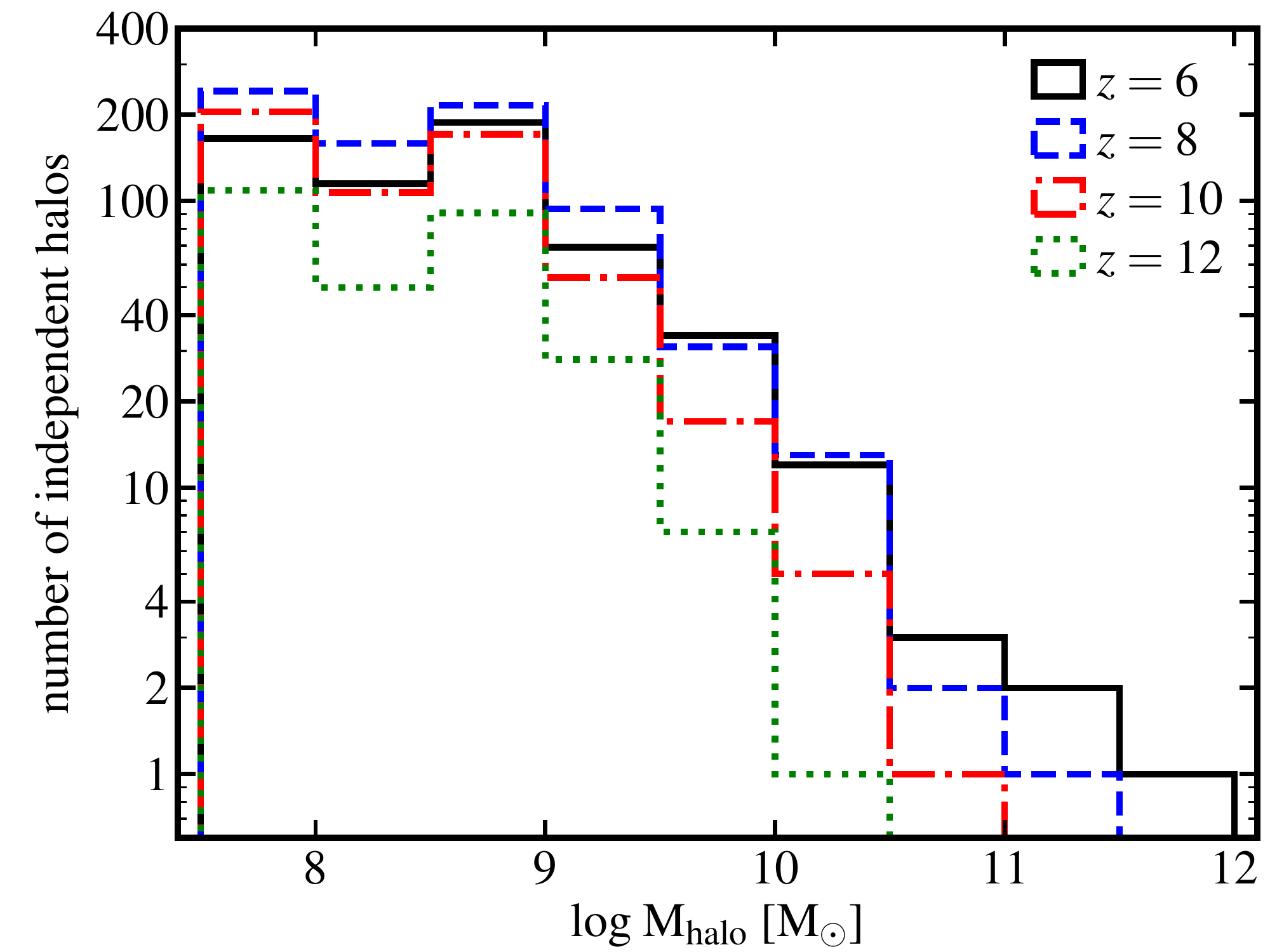} 
\caption{Number of independent halos in the simulated catalog at several redshifts. The simulated sample includes considerably larger numbers of independent halos below $\Mhalo=10^{11}\,\Msun$ at $z\sim6$ or below $\Mhalo=10^{10.5}\,\Msun$ at $z\sim10$, where we are able to account for (at least some) halo-to-halo variance. At higher masses, our sample is limited. We can, however, study time-variability and its impact on scatter in observational properties of galaxies.}
\label{fig:nhalo}
\end{figure} %

\subsection{Baryonic physics}
\label{sec:physics}
We briefly review the baryonic physics here, but refer to \citet[][{sections 2.3--2.5 and appendix B--E}]{hopkins.2017:fire2.numerics} for details. In the simulations, gas follows an ionized-atomic-molecular cooling curve from $10$--$10^{10}$\,K, including metallicity-dependent fine-structure and molecular cooling at low temperatures and high-temperature metal-line cooling for 11 separately tracked species \citep[H, He, C, N, O, Ne, Mg, Si, S, Ca, and Fe; see][]{wiersma.2009:metal.cooling}. We do not include a primordial chemistry network nor Pop III star formation, but apply a metallicity floor of $Z=10^{-4}\,\Zsun$, which corresponds crudely to the metallicity expected after enrichment by the first supernovae (SNe) from Pop III stars \citep[e.g.][]{bromm.2003:first.supernova,wise.2012:primordial.metal.enrich}. This is a reasonable treatment since we mainly focus on relatively massive galaxies at $z\lesssim15$, which are dominated by Pop II stars. At each timestep, the ionization states and cooling rates H and He are calculated following \citet{katz.1996:sph.cooling}, and cooling rates from heavier elements are computed from a compilation of {\sc cloudy} runs \citep{ferland.2013:cloudy.release}, applying a uniform but redshift-dependent photo-ionizing background from \citet{cafg.2009:uvb}\footnote{The ionizing background starts at $z=10.6$, with the ionization rate and heating rate increasing with time until the simulations end at $z=5$. We note that both rates show a sharp increase just below $z\sim7$. A tabulated version of the background is available at http://galaxies.northwestern.edu/uvb/.}, and an approximate model for H\,{\sc ii} regions generated by local sources. Gas self-shielding is accounted for with a local Jeans-length approximation, which is consistent with the radiative transfer calculations in \citet{cafg.2010:lya.cooling}. The on-the-fly calculation of ionization states is broadly consistent with more accurate post-processing radiative transfer calculations \citep{ma.2015:fire.escfrac}.

We follow the star formation criteria in \citet{hopkins.2013:sf.criteria} and allow star formation to take place only in dense, molecular, and locally self-gravitating regions with hydrogen number density above a threshold $\nc=1000\,\cm^{-3}$. Stars form at 100\% efficiency per free-fall time when the gas meets these criteria, and there is no star formation elsewhere. Note that star-forming particles can reach densities much higher than $\nc$ following the self-gravitating criterion. The simulations include several different stellar feedback mechanisms, including (1) local and long-range momentum flux from radiative pressure, (2) energy, momentum, mass and metal injection from SNe and stellar winds, and (3) photo-ionization and photo-electric heating. Every star particle is treated as a single stellar population with known mass, age, and metallicity, assuming a \citet{kroupa.2002:imf} IMF from 0.1--$100\,\Msun$. All feedback quantities are directly calculated from {\sc starburst99} \citep{leitherer.1999:sb99}. 

Note that {\sc starburst99} is a single-star stellar population model\footnote{Note that the stellar population models used in the simulations do not including stellar rotation, which is another key ingredient in stellar population synthesis \citep[e.g.][]{leitherer.2014:rotation.sb99,choi.2017:rotate.massive.star.pop} and could have important implications for reionization \citep[e.g.][]{topping.shull.2015:rotation.ion}.}, which assumes each star evolves independently, but most massive stars are expected to interact with a companion during their lifetimes. This will have significant effects on the SED of young populations, especially at low metallicities \citep[e.g.][]{stanway.2017:massive.star.iau.proc}. It has been suggested that massive binaries can lead to high escape fractions of ionizing photons from high-redshift metal-poor galaxies, and thus have important implications for understanding the sources dominating cosmic reionization \citep{ma.2016:fire.fesc.binary,gotberg.2017:binary.ionizing.spec}. Nonetheless, binarity only has weak effects on most stellar feedback quantities, such as bolometric luminosities (within 0.05\,dex in the first 200\,Myr since a stellar population is born) and Type-II SNe rates \citep[e.g.][]{xiao.eldridge.2015:bpass.sne.rate}, so we do not expect binary interaction to have significant dynamical effects\footnote{Binary models do produce more ionizing photons (see Section \ref{sec:bpass}), which are likely to enhance photo-ionization feedback, but we checked that this only has sub-dominant effects on gas dynamics \citep{ma.2015:fire.escfrac}.}. For these reasons, we only consider binary stellar population models in post-processing. In this paper, we use the Binary Population and Spectral Synthesis (BPASS) models \citep[version 2.0;][]{eldridge.2008:bpass.model,stanway.2016:binary.ion.budget}\footnote{http://bpass.auckland.ac.nz} to compute the SED of each star particle from its age and metallicity. The BPASS models include both single-stellar and binary stellar population synthesis models. Their single-star models agree well with {\sc starburst99}. Their binary models take into account mass transfer, common envelope phase, binary mergers, and quasi-homogeneous evolution at low metallicities. Also, the BPASS binary models appear to explain the nebular emission line properties observed in $z\sim2$--3 galaxies \citep[e.g.][]{steidel.2016:stellar.nebular.spec,strom.2017:nebular.line.mosfire}. In this paper, we mainly consider stellar continuum emission, while detailed modeling of dust extinction and nebular line emission will be the subject of future studies.

\subsection{Halo selection and definitions}
\label{sec:definition}
We run {\sc ahf} on every snapshot to identify halos and subhalos in the zoom-in region. In general, most stars of the central (satellite) galaxy in a halo (subhalo) reside in $\frac{1}{3}\Rmax$ from the halo center, where $\Rmax$ is the radius at which the maximum circular velocity $\Vmax$ is reached ($\Rmax$ is already computed by {\sc ahf}). We thus define a galaxy by including all star particles within $\frac{1}{3}\Rmax$ after removing contributions from subhalos outside $\frac{1}{5}\Rmax$. This excludes star particles at large distances from the halo center (corresponding to diffuse stellar distributions) and allow us to mask satellites. For each galaxy, we obtain a list of star particles with known position, age, and metallicity, from which we can compute a number of galaxy properties, such as stellar mass, star formation history, broad-band luminosities and magnitudes, surface brightness, galaxy size, etc. In this paper, we primarily focus on central galaxies, which dominate the stellar light: at a given stellar mass, only a few per cent of the galaxy population are satellites. We have also confirmed that they do not differ significantly from centrals at similar stellar masses in most properties we study in this paper. We restrict our analysis below to {central} halos that have zero contamination from low-resolution particles within $\Rvir$ and contain more than $10^4$ particles in total\footnote{{This excludes most halos below $\Mhalo\sim10^{8.6}$ ($10^{7.7}$) $\Msun$ in simulations at resolution $m_b\sim7000$ (900) $\Msun$. The minimum number of dark matter particles for halos in our simulated catalog is $\sim5600$.}}. Our target halos are guaranteed to meet this criteria by construction, but we also consider other halos in the zoom-in regions in our analysis.

In Figure \ref{fig:nhalo}, we show the number of halos that meet the above criteria in all zoom-in regions at several redshifts. Our simulations are able to capture (at least some) halo-to-halo variance below $\Mhalo=10^{11}\,\Msun$ at $z\sim6$ and below $\Mhalo=10^{10.5}\,\Msun$ at $z\sim10$, where the simulations include a few to more than 200 halos in a given halo mass bin. Moreover, these galaxies always have `bursty' star formation histories (see Section \ref{sec:sfr}), which leads to significant time variability in their properties \citep[e.g.][]{muratov.2015:fire.mass.loading,sparre.2017:fire.sf.burst,feldmann.2017:massive.fire.long,ma.2017:fire.metallicity.gradient,cafg.2018:bursty.sf.model}. Hence a galaxy tends to move above and below the median of certain scaling relations (see also the discussion in Section \ref{sec:smhm}). To account for the scatter due to bursty star formation (as well as galaxy mergers and other time-variable phenomena), we make use of 58 snapshots saved for each simulation at redshifts $z=5$--12 (about 20\,Myr apart between two successive snapshots) to build a catalog of over 34,000 simulated halo `snapshots'. By doing so, we sample the same halos multiple times in the catalog and treat them as statistically equal in our analysis. Figure \ref{fig:nhalo} essentially shows the number of {\em independent} halos in the simulated sample. At lower masses (e.g. $\Mhalo\leq10^{11}\,\Msun$ at $z\sim6$), we are able to account for the scatter both from halo-to-halo variance and time variability within single halos. A priori, it is not clear which effect dominates the scatter for a given scaling relation. At higher masses ($\Mhalo\geq10^{11}\,\Msun$), our sample only contains 1--2 independent halos at a given redshift, so we are only able to account for the variance due to time variability of individual galaxies. We caution that we may therefore underestimate the scatter of certain scaling relations at the high-mass end. We have also checked that excluding a randomly selected 1/2--2/3 of the snapshots from our analysis (sampling each galaxy at sparser time steps) does not change the results of this paper. In other words, our time-sampling is sufficient for statistically converged results.

\begin{figure*}
\centering
\begin{tabular}{cc}
$u/g/r$-composite images ($z=5$) & 
Mock JWST NIRCam F277W images ($z=5$) \\
\includegraphics[width=0.48\linewidth]{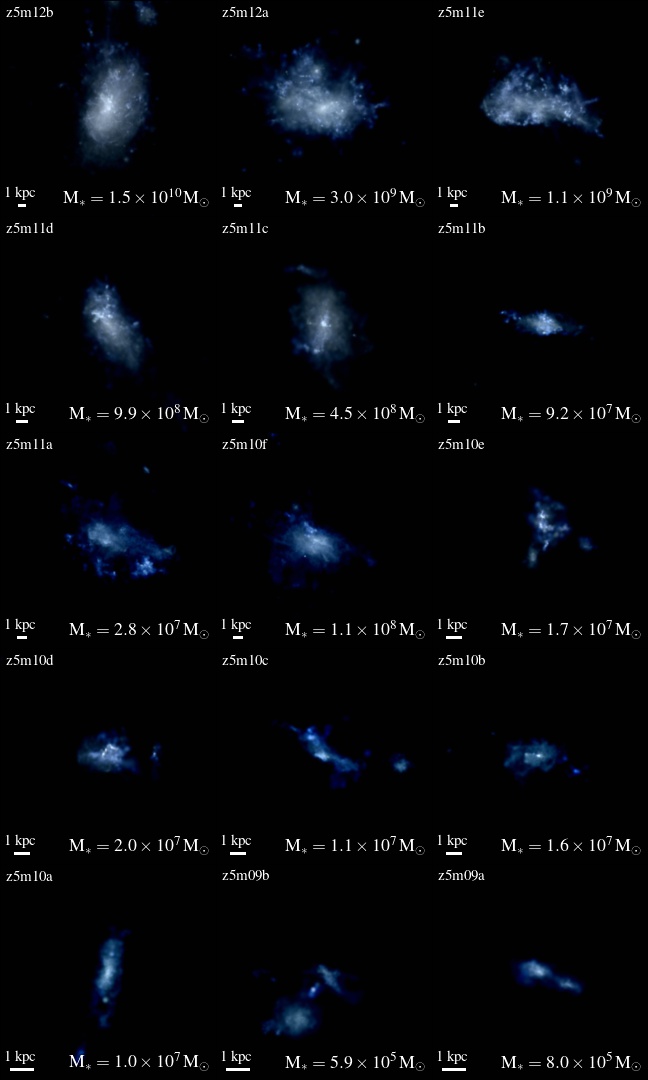} &
\includegraphics[width=0.48\linewidth]{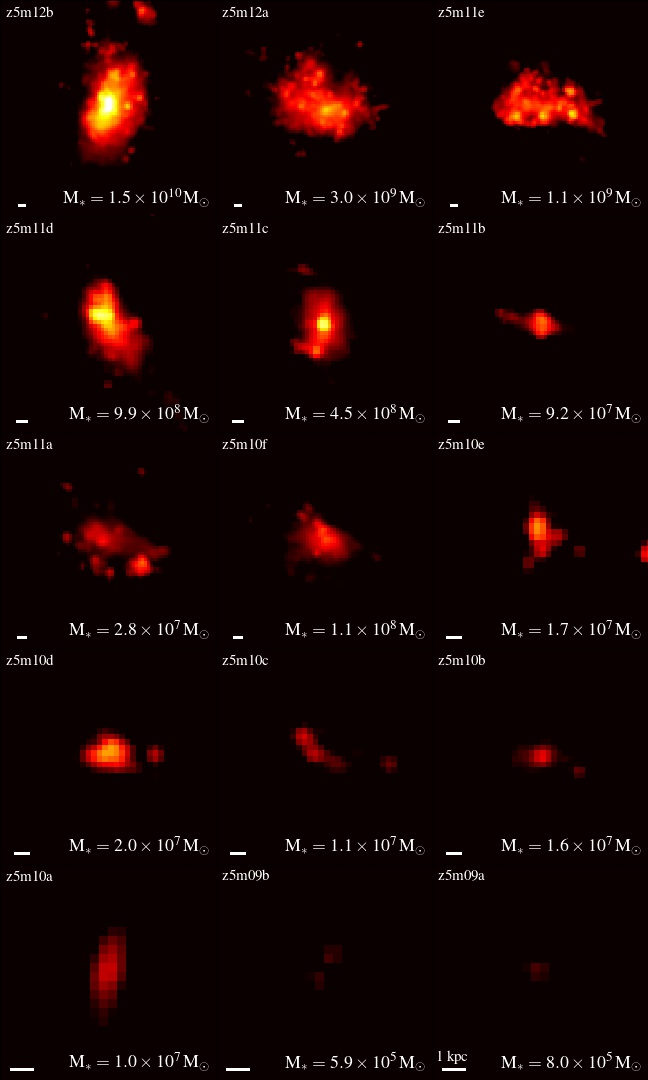} \\
\end{tabular}
\caption{{\em Left:} Stellar $u/g/r$-composite images for the central galaxies of the `target' halos from Table \ref{tbl:sim} at $z=5$. {\em Right:} Noise-free mock JWST NIRCam F277W-band images (rest-frame 4600\,\AA). The PSF is a Gaussian function with FWHM of 2 pixels. The pixel size is 0.065\,arcsec and 0.42\,kpc in physical length. The three images in the same row use the same color scale, which spans eight magnitudes in surface brightness, but the depth increases from $m_{\rm AB}=29.5$\,mag\,arcsec$^{-2}$ for the most massive galaxies in the top row to 31.5\,mag\,arcsec$^{-2}$ for low-mass galaxies in the bottom row. We use the BPASSv2.0 binary models to determine the SED of each star particle from its age and metallicity, and then ray trace along the line-of-sight without dust attenuation. Nebular line emission is also ignored. The scale bar in each panel indicates 1\,kpc (physical).}
\label{fig:imz5}
\end{figure*} %

\begin{figure*}
\centering
\begin{tabular}{cc}
$u/g/r$-composite images ($z=10$) & 
Mock JWST NIRCam F444W images ($z=10$) \\
\includegraphics[width=0.48\linewidth]{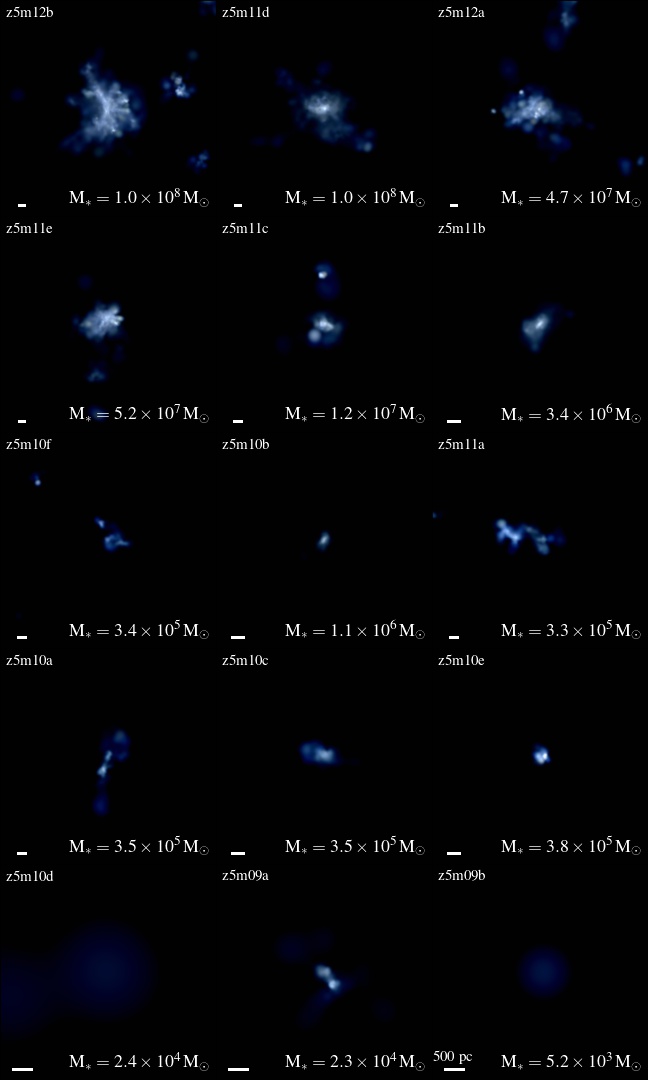} &
\includegraphics[width=0.48\linewidth]{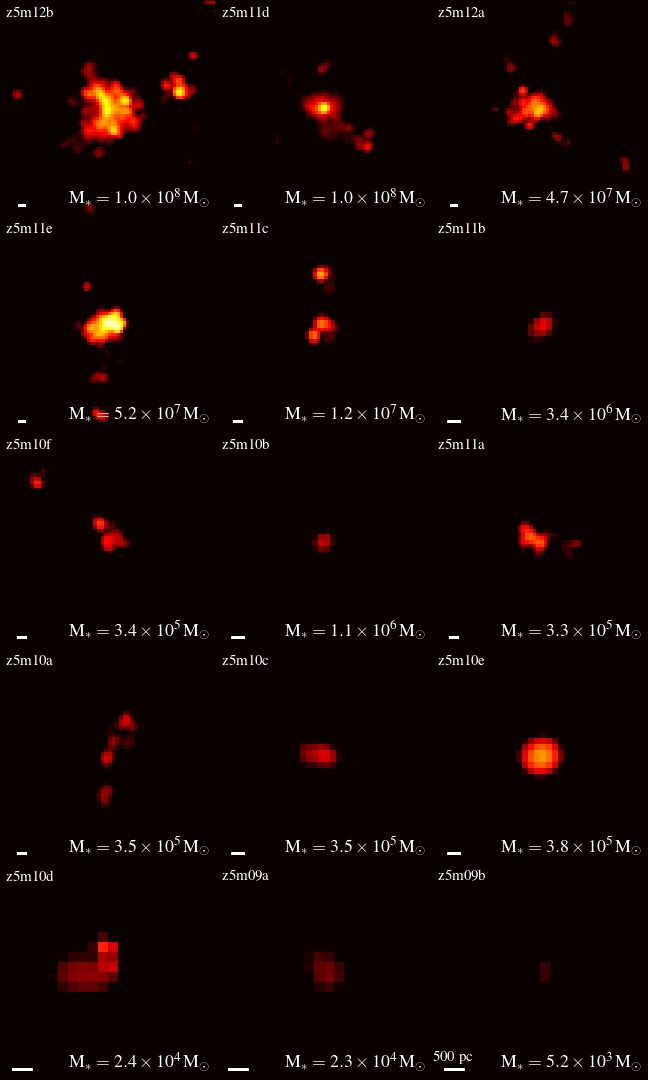} \\
\end{tabular}
\caption{{\em Left:} Stellar $u/g/r$-composite images for all central galaxies at $z=10$ (as Figure \ref{fig:imz5}). {\em Right:} Noise-free mock JWST NIRCam F444W-band images (rest-frame 4000\,\AA). The PSF is a Gaussian function with FWHM of 2 pixels. The pixel size is 0.065\,arcsec and 0.28\,kpc in physical length. The three images in the same row use the same color scale, which spans eight magnitudes in surface brightness, but the depth increases from $m_{\rm AB}=32$\,mag\,arcsec$^{-2}$ for the most massive galaxies in the top row to 36\,mag\,arcsec$^{-2}$ for low-mass galaxies in the bottom row. The scale bar in each panel indicates 500\,pc (physical).}
\label{fig:imz10}
\end{figure*} %

\begin{figure*}
\centering
\includegraphics[width=0.95\textwidth]{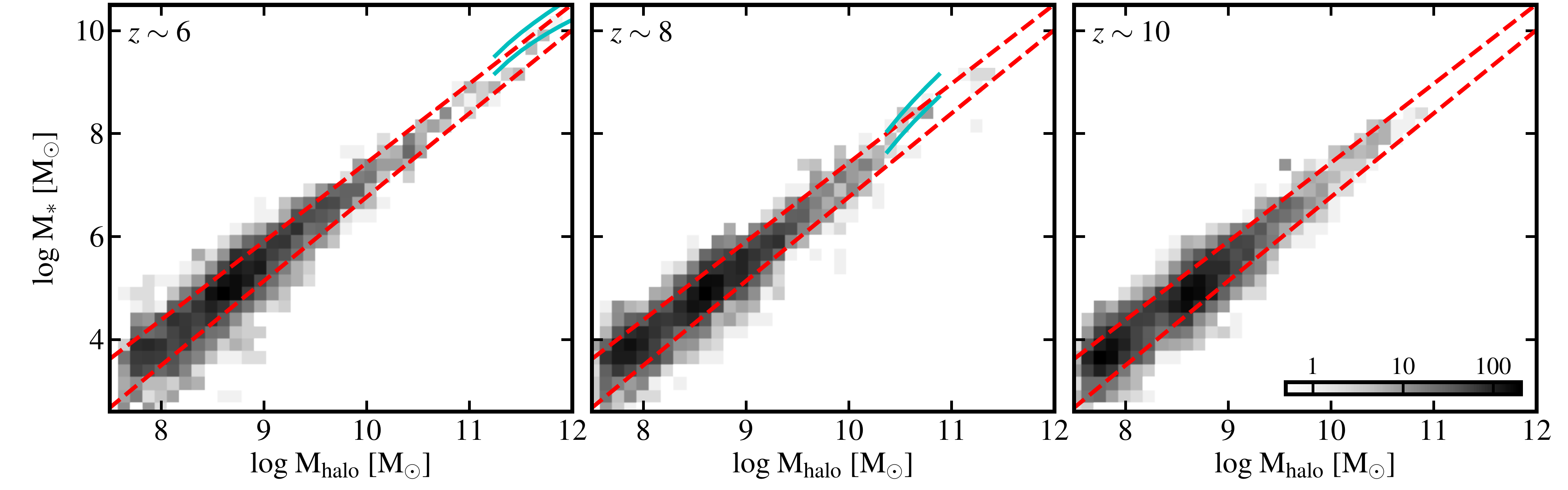} \\
\includegraphics[width=0.95\textwidth]{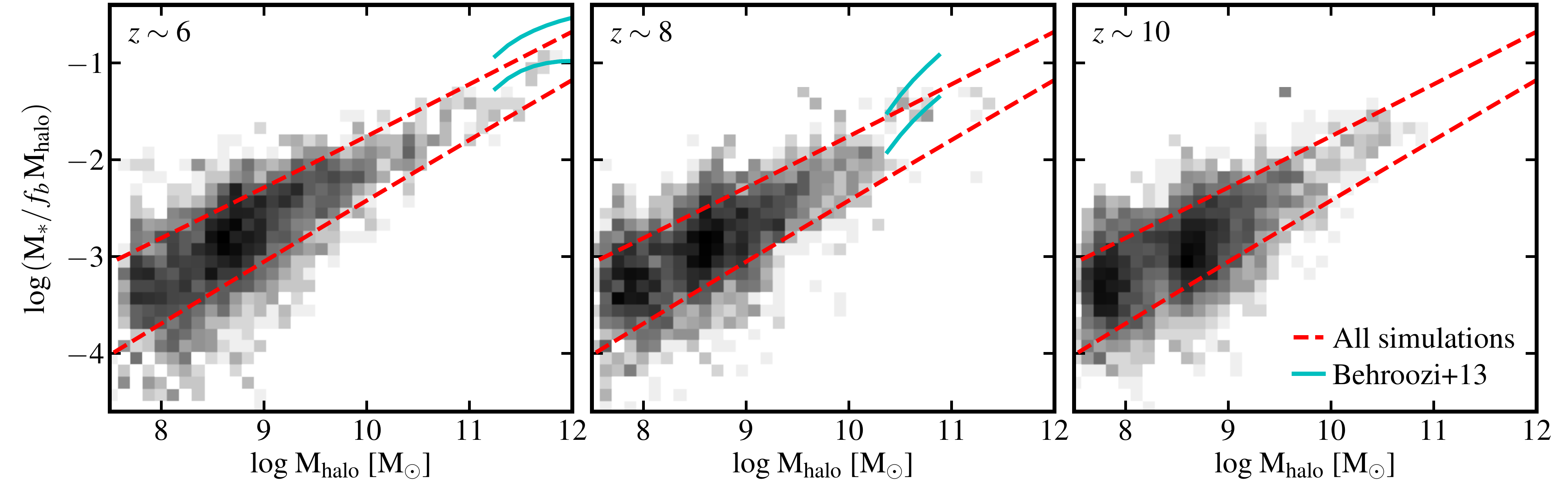} 
\caption{{\em Top:} The stellar mass--halo mass relation at $z=6$, 8, and 10. {\em Bottom:} The stellar baryon fraction--halo mass relation at the same redshifts. The two-dimensional histograms represent the number of simulated halo snapshots in each pixel in logarithmic scale (as shown by the color scale). All central galaxies that meet the selection criteria described in Section \ref{sec:definition} are included. The cyan lines show the abundance matching results at $z=5$--8 from \citet{behroozi.2013:abundance.matching}. The red dashed lines show the best-fit $1\sigma$ region of {\em all} central galaxies at $z=5$--12 (lines are identical in all three panels, see Section \ref{sec:smhm} for details). At each redshift, the stellar mass correlates tightly with halo mass, and there is no significant evolution in the stellar mass--halo mass relation from $z=5$--12.}
\label{fig:smhm}
\end{figure*} %

\subsection{Halo abundances}
\label{sec:weights}
Since our simulated catalog is constructed from 15 cosmological zoom-in regions, it does not contain information about the halo abundance at a given halo mass and redshift. Therefore, we assign every simulated halo `snapshot' a weight to recover the appropriate number density of halos at its mass and redshift in the Universe. {We briefly summarize the method here and refer the readers to Appendix \ref{app:weight} for details.} We use {\tt HMFcalc} \citep{murray.2013:hmfcalc}\footnote{http://hmf.icrar.org} to calculate the halo mass functions, applying the same cosmological parameters and virial overdensities as those adopted in the simulations. We take the fitting functions from \citet{behroozi.2013:abundance.matching} in {\tt HMFcalc}, which is a modified \citet{tinker.2008:halo.mass.function} halo mass function. It matches well with the halo mass functions directly extracted from our large-volume dark matter-only cosmological boxes in the redshift range we consider here. We bin the simulated sample in the two-dimensional $\log\Mhalo$--$\log(1+z)$ space with bin widths $\Delta\log\Mhalo=0.4$ from $\Mhalo=10^{7.5}$--$10^{12}\,\Msun$ and $\Delta\log(1+z)=0.04$ from $z=5$--12. We have confirmed that our results are not sensitive to the bin widths we adopt. In each bin, we count the number of halos in the simulated catalog $N_{\rm sim}$ and calculate the number of halos expected in the Universe $N_{\rm expect}$ from the halo mass function. All halos in the same bin are assigned the same weight $w=N_{\rm expect}/N_{\rm sim}$. In other words, by summing $w$ over all simulated halos in certain halo mass and redshift intervals and dividing $\sum_i w_i$ by the corresponding comoving volume, one should recover the halo number densities given by the input halo mass functions. When necessary, each halo snapshot in the simulated catalog is weighted by its $w$. This is important when we consider statistical properties of simulated galaxies at a fixed stellar mass or magnitude, where not all galaxies have equal halo mass (e.g. Sections \ref{sec:sfr} and \ref{sec:mag}). The weights will also be used to construct stellar mass functions and luminosity functions in Sections \ref{sec:smf} and \ref{sec:lf}.

\referee{Strictly speaking, this approach is valid only if the halos in our samples are not strongly biased. However, after an extensive check, we find no significant difference between halos in various environments \citep[see also e.g.][]{oshea.2015:renaissance.uvlf} and at different resolution in our simulations regarding their properties studied in this paper. Although our sample is still possibly biased due to complex selection criteria -- for example, all halos below $10^9\,\Msun$ by $z=5$ in our sample live within a few virial radii of a more massive halo (i.e., the target halo in the zoom-in region) and we lack isolated halos at such low masses down to $z=5$, our conclusions in this paper are likely robust.}

\section{Galaxies in the reionization era}
\label{sec:galaxies}

\subsection{Morphology}
\label{sec:morphology}
In Figure \ref{fig:imz5}, we show the stellar $u/g/r$-composite images at $z=5$ (left) for the central galaxy in the most massive halo in each zoom-in region. The stellar masses and halo masses are listed in Table \ref{tbl:sim}. We use the BPASSv2.0 binary models to determine the stellar SEDs, assuming a \citet{kroupa.2002:imf} IMF from 0.1--$100\,\Msun$. Note that we only consider intrinsic stellar continuum emission, and ignore dust extinction and nebular line emission at this point\footnote{{Full spectral modeling of high-redshift galaxies in cosmological simulations has been developed recently by other groups \citep[e.g.][]{wilkins.2016:bluetides.photometry,barrow.2017:highz.galaxy.spectra,zackrisson.2017:spectral.first.gal}.}}. The right panel shows the noise-free mock images as observed by the Near Infrared Camera (NIRCam) on JWST at F277W band (rest-frame 4600\,\AA), applying a Gaussian point spread function (PSF) with full width half maximum (FWHM) of two pixels with pixel size 0.065\,arcsec (0.42\,kpc in physical length)\footnote{The PSF and pixel sizes are adopted from the NIRCam pocket guide from https://jwst.stsci.edu/instrumentation/nircam.}. The three images in the same row are shown using the same color scale, which spans eight magnitudes in surface brightness, but the depth increases from $m_{\rm AB}=29.5$\,mag\,arcsec$^{-2}$ in the top row to 31.5\,mag\,arcsec$^{-2}$ in the bottom row (pixels below these limits are shown as black).

In Figure \ref{fig:imz10}, we show the stellar $u/g/r$-composite images and noise-free mock JWST NIRCam F444W-band images (rest-frame 4000\,\AA) at $z=10$ for the most massive galaxy in each zoom-in simulation. These images are rearranged in place to ensure a descending order in halo mass from the top-left panel to the bottom-right panel. The mock JWST images have a pixel size 0.065\,arcsec and 0.28\,kpc in physical length. Again, the color scale in each image spans eight magnitudes in surface brightness, but the depth increases from $m_{\rm AB}=32$\,mag\,arcsec$^{-2}$ in the top row to 36\,mag\,arcsec$^{-2}$ in the bottom row.

Almost all of the simulated galaxies at $z\geq5$ show clumpy, irregular morphologies even in rest-frame optical bands, possibly due to high merger rates and clumpy, gas-rich star formation at these redshifts. This is in contrast to galaxies at low and intermediate redshifts, which show a mix of late-type, early-type, and irregular morphologies at similar masses \citep[e.g.][]{feldmann.2017:massive.fire.long,elbadry.2018:gas.kinematics}. Galaxies with similar stellar mass may have a variety of sizes and surface brightness, so their detectability can differ significantly. Therefore, our high-resolution simulations provide a useful database for understanding future multi-band, spatially resolved observations of $z\gtrsim5$ galaxies, as well as determining the completeness of a flux-limited galaxy survey at these redshifts.

\subsection{The stellar mass--halo mass relation}
\label{sec:smhm}
Figure \ref{fig:smhm} shows the stellar mass--halo mass relation (top panels) and the stellar baryon fraction--halo mass relation (bottom panels) for central galaxies at $z=6$, 8, and 10. The stellar baryon fraction is defined as $\Ms/(f_b\Mhalo)$, where $f_b=\Omega_b/\Omega_m$ is the cosmic baryonic fraction. The two-dimensional histograms represent the number of halo snapshots in the simulated catalog (as defined in Section \ref{sec:definition}) within $\Delta z=0.5$ (e.g. from $z=7.5$--8.5 in the $z=8$ panels) in each pixel in logarithmic scale (as shown by the color scale). We remind the readers that we re-sample each halo multiple times to account for time-variability in where galaxies lie on this relation, but we refer to Figure \ref{fig:nhalo} for the number of independent halos in our sample at these redshifts (see Section \ref{sec:definition} for details). We also show the empirical relations at $z=6$ and 8 from \citet[][the cyan lines]{behroozi.2013:abundance.matching}. At all redshifts, there is a tight correlation between stellar mass and halo mass, with the scatter increasing at the low-mass end\footnote{We caution that our simulated sample have more independent galaxies at low masses than at higher masses, so we may underestimate the scatter at the high-mass end. Nonetheless, a halo mass-dependent scatter in the stellar mass--halo mass relation does exist at low redshift in both observations and FIRE-2 simulations \citep[e.g.][]{sgk.2014:elvis.abundance.match,fitts.2017:field.dwarf.galaxy}. The current simulations are consistent with increased scatter at low masses.}. We also examine the relation for satellite galaxies (not shown), which tend to have systematically higher stellar mass and larger scatter than central galaxies at a given halo mass, due to the fact that their halos are usually stripped. However, we note that the halo mass of a satellite depends strongly on which halo finder one uses. We find a smaller offset between central and satellite galaxies using the {\sc rockstar} subhalo catalog than using the {\sc ahf} catalog. Because satellite galaxies contribute no more than a few per cent of the total galaxy population at a given mass, we do not further quantify the difference in this paper.

We find little evolution in the stellar mass--halo mass relation at these redshifts, \referee{in line with recent empirical constraints (e.g. \citealt{rodriguez-puebla.2017:new.smhm}; however, see \citealt{behroozi.silk.2015:hiz.gal.model})}. We will show it explicitly below. Using all halo snapshots in the simulated catalog at $z=5$--12, we calculate the median and $1\sigma$ dispersion in $\log\Ms$ at every 0.5\,dex in $\log\Mhalo$ from $\Mhalo=10^{7.5}$--$10^{12}\,\Msun$. We assume a simple power-law relation between $\Ms$ and $\Mhalo$
\be
\label{eqn:smhm}
\log\Ms = \alpha \, (\log\Mhalo - 10) + \beta
\ee
and a halo mass-dependent scatter
\be
\label{eqn:scatter}
\sigma_{\log\Ms} = \exp \, [\gamma \, (\log\Mhalo - 10) + \delta],
\ee
and fit the median stellar mass--halo mass relation and $1\sigma$ dispersion obtained from the simulated sample as described above. We obtain the best-fit parameters as 
\be
(\alpha,\,\beta,\,\gamma,\,\delta)=(1.58,\,7.10,\,-0.14,\,-1.10). 
\ee
These results are nearly identical to those obtained from the FIRE-1 simulations \citep[e.g.][]{hopkins.2014:fire.galaxy,ma.2015:fire.escfrac,feldmann.2017:massive.fire.long} at similar halo mass and redshift, despite the subtle differences in numerical details and resolution between these simulations. \referee{Our predictions also broadly agree with those in the literature \citep[e.g.][]{ceverino.2017:firstlight.project}.} We show our best-fit $1\sigma$ stellar mass--halo mass relation in every panel in Figure \ref{fig:smhm} (the red dashed lines). Visual inspection implies that equations \ref{eqn:smhm} and \ref{eqn:scatter} describe our simulated sample reasonably well at any redshift. We also confirmed that the median relation obtained from a subsample at a given redshift does not deviate from Equation \ref{eqn:smhm} by more than 0.1\,dex at most halo masses we consider here. It is an intriguing question why the $\Ms$--$\Mhalo$ relation does not evolve at these redshifts. We speculate that this is probably due to feedback regulating star formation to zeroth order. A detailed analysis of the relation between halo growth rate, gas accretion rate, and stellar mass growth rate to understand the weak-evolution of the $\Ms$--$\Mhalo$ relation is worth future investigation.

\referee{}

\begin{figure*}
\centering
\includegraphics[width=\linewidth]{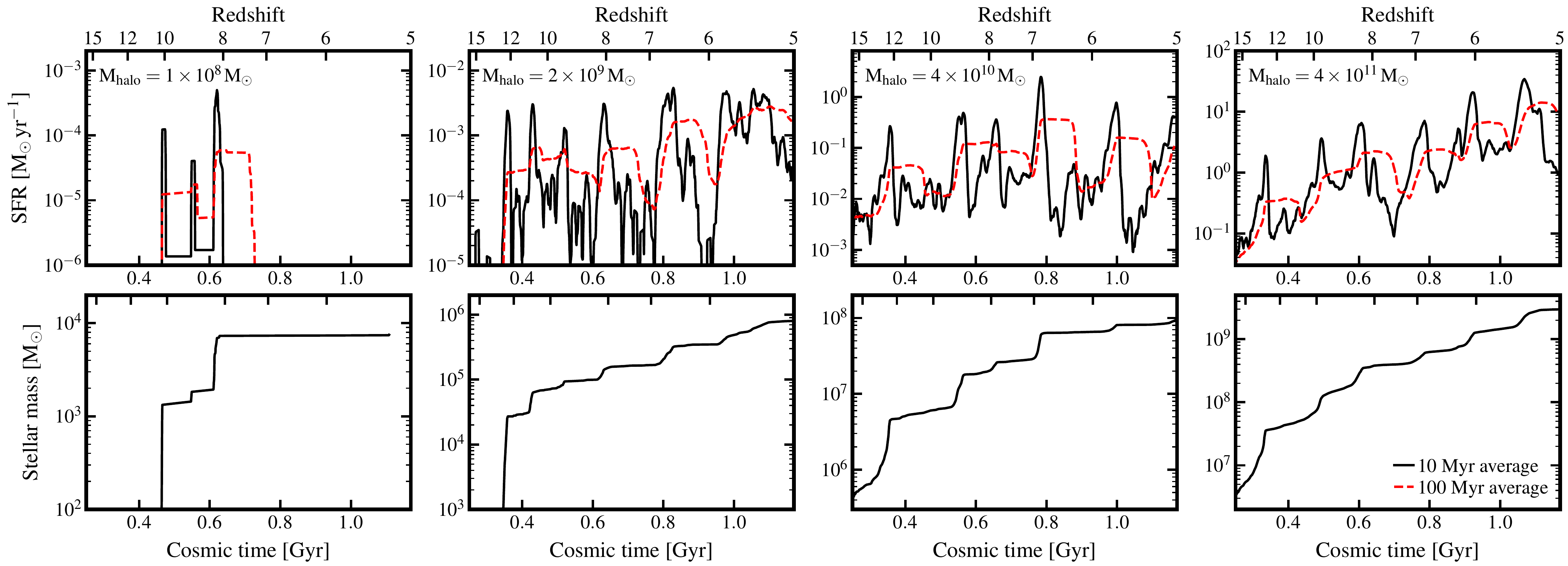} 
\caption{{\em Top:} Star formation histories averaged on 10\,Myr time-scale (black solid lines) and 100\,Myr time-scale (red dashed lines) for four example galaxies. The $z=5$ halo masses are labeled in the upper-left corner. {\em Bottom:} Stellar mass growth histories for the same galaxies. All high-redshift galaxies show strong `bursty' star formation histories. The least massive halo ($\Mhalo=10^8\,\Msun$, left panels) does not form any stars after $z\sim8$: the feedback following a starburst removes the majority of its gas at that time, and the ionizing background prevents fresh gas from accreting and cooling efficiently onto such low-mass halos at late times. More massive halos remain star forming until the end of the simulation at $z=5$.}
\label{fig:sfh}
\end{figure*} %

How do galaxies evolve on the $\Ms$--$\Mhalo$ relation? All of our simulated galaxies experience bursty star formation because of stellar feedback. The stellar mass can grow by a factor of 2 or more during a short time period at the peak of a starburst, while it can remain almost unchanged during the troughs of its star formation history (see Section \ref{sec:sfr} and Figure \ref{fig:sfh} for examples). In contrast, the halo mass grows relatively smoothly via dark matter accretion, which is less affected by feedback. As a consequence, a galaxy moves vertically on the $\Ms$--$\Mhalo$ plane during the peak of a starburst and reaches some point above the median $\Ms$--$\Mhalo$ relation, while it then moves horizontally during a trough in its star formation history and reaches somewhere below the median relation until the next starburst episode. We confirmed in our simulations that the scatter in the $\Ms$--$\Mhalo$ relation caused by bursty star formation is a physical effect. 

There are several caveats in this analysis. First, at the high-mass end ($\Mhalo\geq10^{11}\,\Msun$), our approach only captures the scatter due to bursty star formation, but the sample does not contain sufficient numbers of independent halos to account for halo-to-halo variance. Therefore, we may underestimate the scatter at these halo masses. At lower masses, our simulations include considerably larger numbers of independent halos and the scatter is reliably measured. Second, our simulations do not include more massive halos above $\Mhalo=10^{12}\,\Msun$. At these masses, slowly-cooling hot halos and feedback from supermassive black holes may play an important role. Studying early galaxy formation in such high-redshift massive halos is beyond the scope of this paper. It may lead to a turnover in the $\Ms$--$\Mhalo$ relation at these redshifts similar to what is seen at lower redshifts \citep[e.g.][]{behroozi.2013:abundance.matching}. We caution that our best-fit $\Ms$--$\Mhalo$ relation may break down at $\Mhalo>10^{12}\,\Msun$. Lastly, the best-fit $\Ms$--$\Mhalo$ relation does not apply to halos with $\Mhalo\lesssim10^8\,\Msun$ below $z\sim6$. We will show in Section \ref{sec:sfr} that star formation in these low-mass halos is suppressed by the ionizing background near the end of reionization.

\begin{figure*}
\centering
\begin{tabular}{cc}
\includegraphics[width=0.48\linewidth]{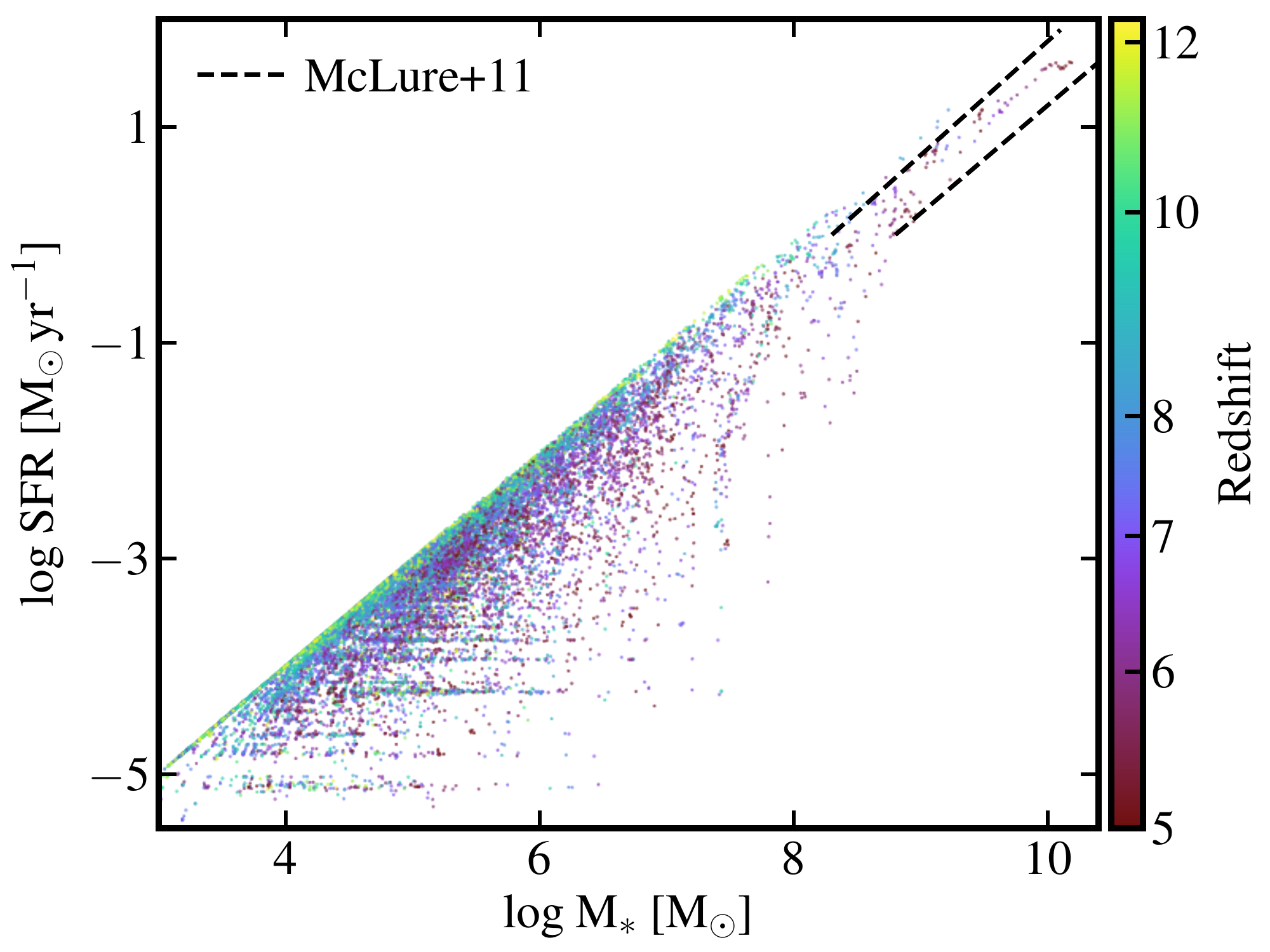} & 
\includegraphics[width=0.48\linewidth]{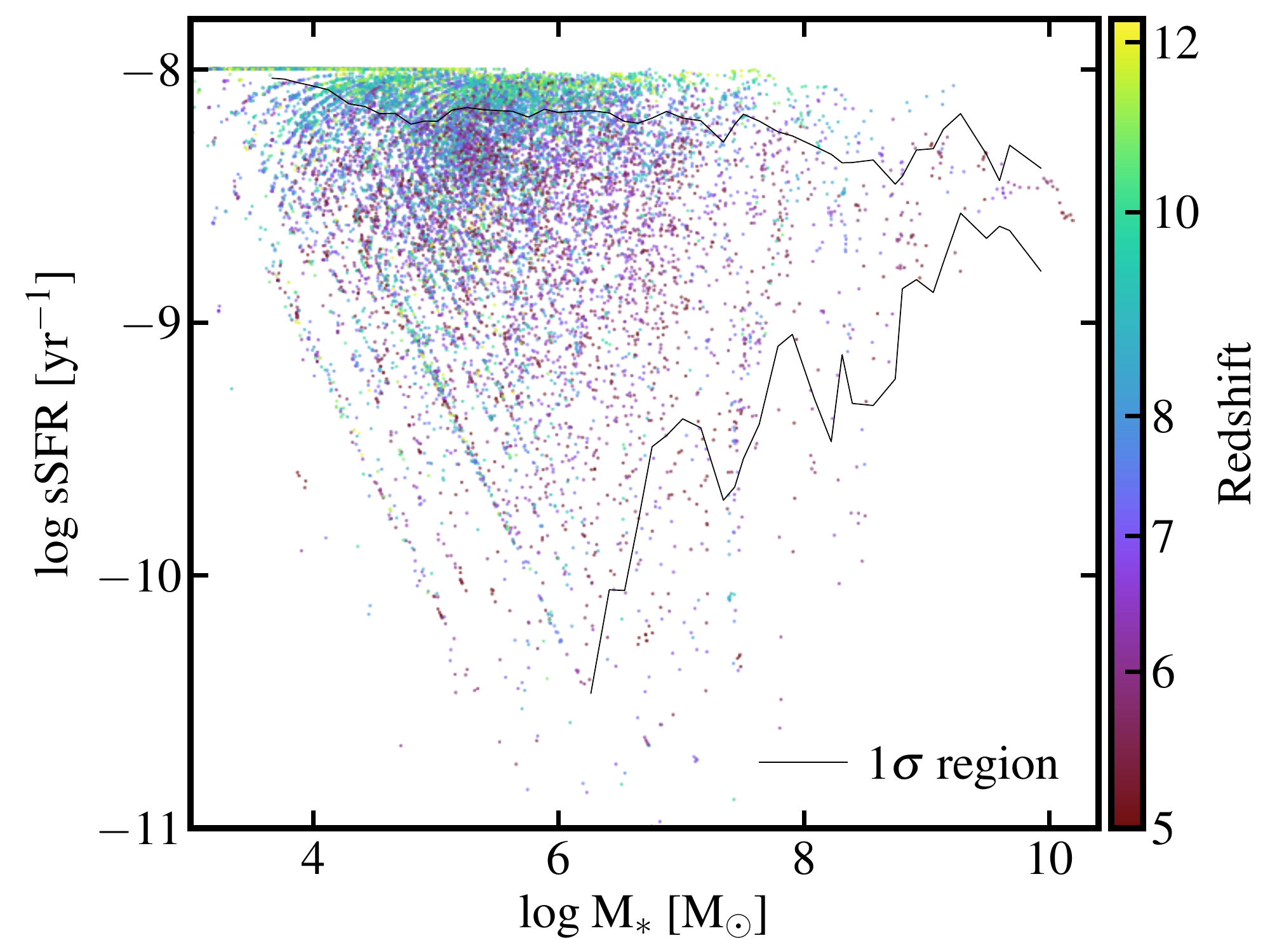} \\
\end{tabular}
\caption{{\em Left:} SFR--$\Ms$ relation. {\em Right:} Specific SFR (sSFR)--$\Ms$ relation. Each point shows a star-forming galaxy snapshot in the simulated sample, color-coded by its redshift. The solid lines in the right panel illustrate the weighted $1\sigma$ region at $z\sim6$. The SFR is averaged over the past 100\,Myr at the time of measurement. The sharp upper limits in both relations are because these galaxies formed nearly all of their stars within the past 100\,Myr. The black dashed lines show the observed relation for $z=6$--9 galaxies from \citet{mclure.2011:highz.galaxy.sfr}. Our simulations agree well with observations at the most massive end. At lower masses, the scatter is large as a result of strong bursty star formation histories. Galaxies at lower redshifts have lower SFRs and larger scatter on average than galaxies at higher redshifts at a given stellar mass.}
\label{fig:sfr}
\end{figure*} %

\begin{figure*}
\centering
\includegraphics[width=\linewidth]{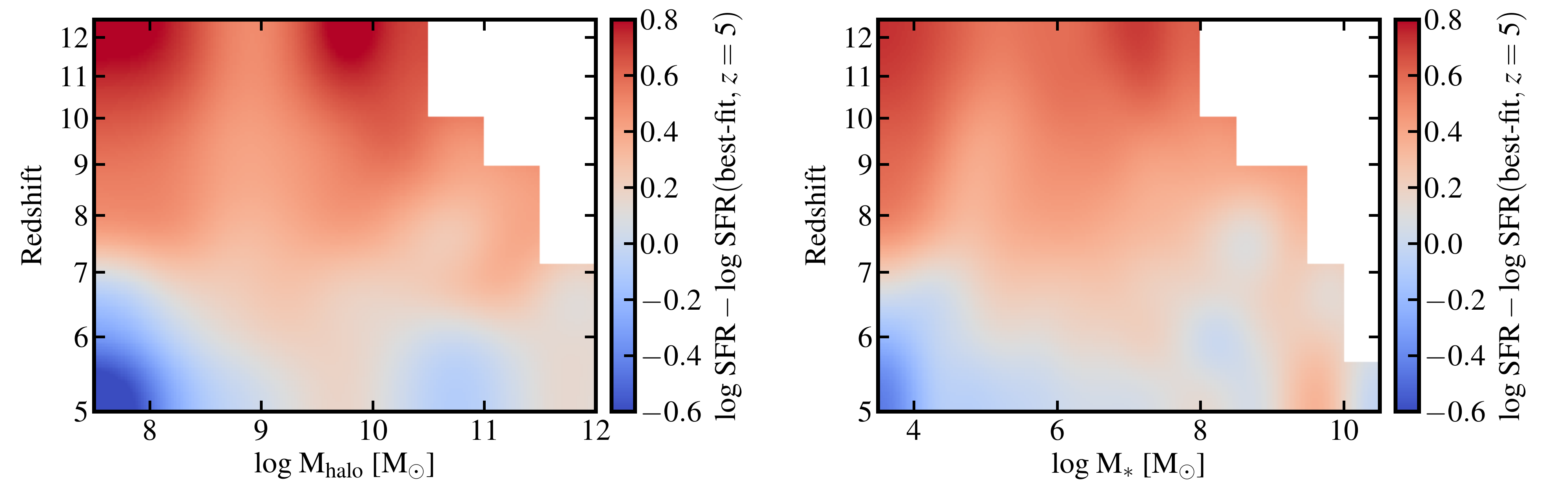} 
\caption{Bivariate relation between SFR, redshift, and halo mass [$\SFR(\Mhalo,\,z)$, left] or stellar mass [$\SFR(\Ms,\,z)$, right]. Colors show the weighted average $\SFR$ for the simulated sample, relative to the best-fit relation at $z=5$ (as labeled beside the colorbars). This emphasizes the dependence on redshift. The white regions represent parameter space where we have no simulations. At fixed halo mass or stellar mass, the mean SFR increases with increasing redshift, by about 0.7\,dex from $z=5$ to $z=12$. Since $z\sim6$, the average SFRs for galaxies below $\Mhalo\sim10^8\,\Msun$ or $\Ms\sim10^4\,\Msun$ decrease by approximately 0.6\,dex, because star formation in low-mass galaxies is suppressed by the ionizing background near the end of reionization.}
\label{fig:sfhz}
\end{figure*} %

\begin{figure*}
\centering
\includegraphics[width=\linewidth]{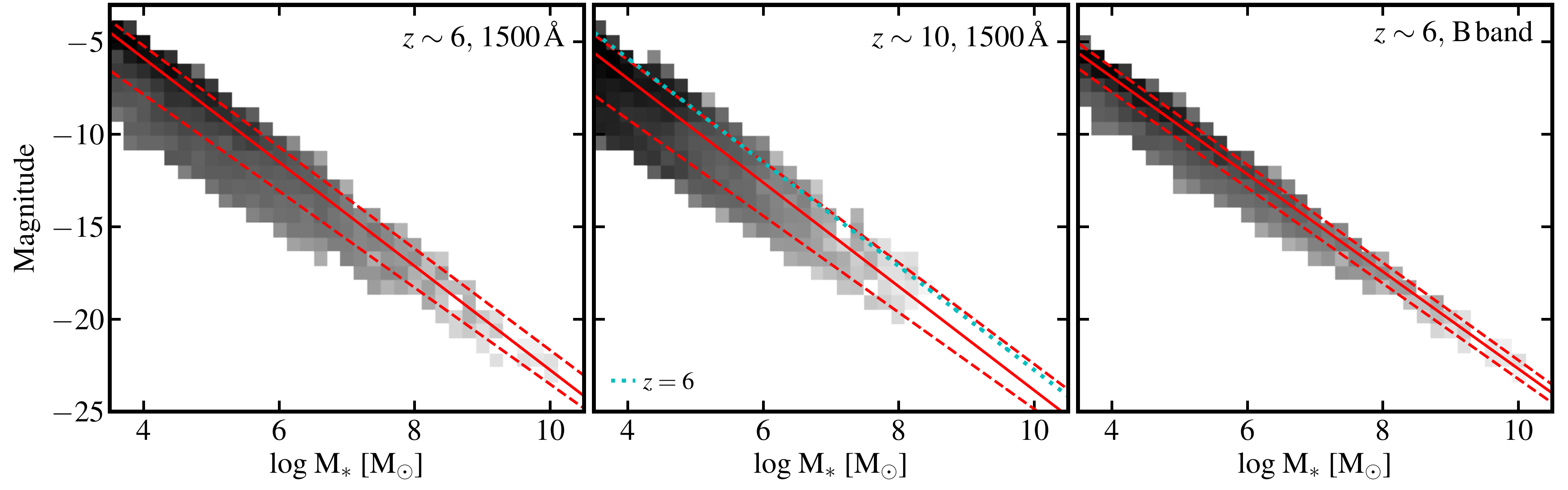}
\includegraphics[width=\linewidth]{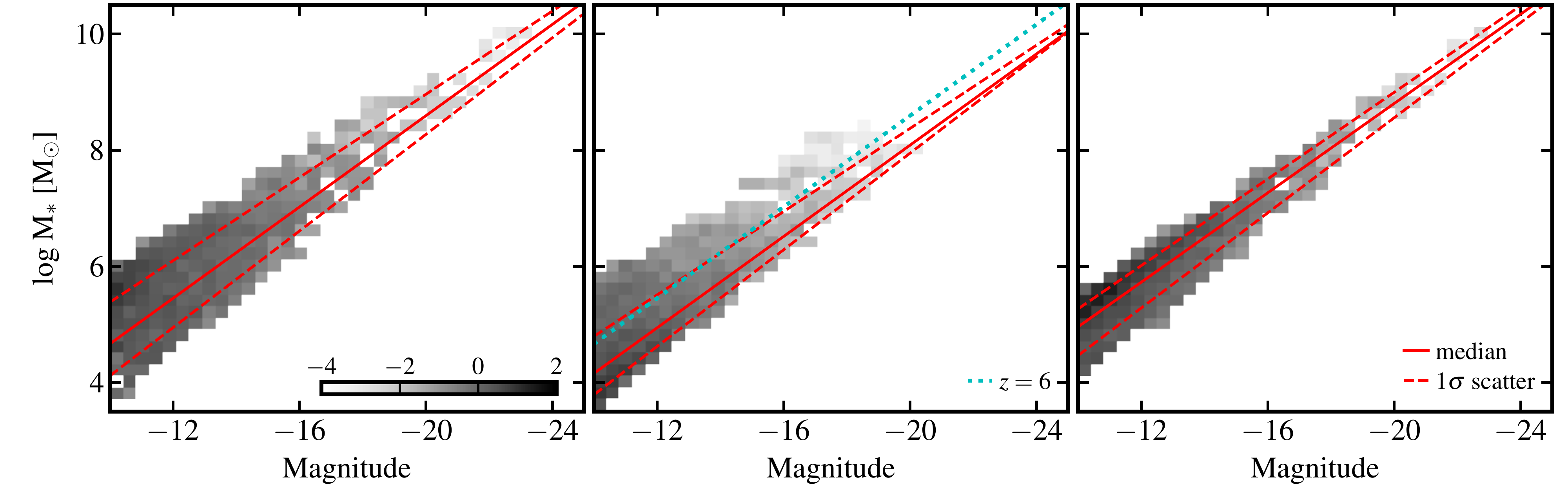}
\caption{{\em Top:} The magnitude--stellar mass relation for rest-frame 1500\,{\AA} at $z=6$ (left) and $z=10$ (middle), and for rest-frame B band at $z=6$ (right). The two-dimensional histogram represents the total {\em weight} of simulated galaxy snapshots in each pixel (in logarithmic scale), taking into account our simulation results and the halo mass function (see Section \ref{sec:weights} for details). This reflects the true relative number of galaxies between pixels in the Universe. The red solid and dashed lines show the best-fit weighted median relation and $1\sigma$ scatter (see text for details). {The cyan dotted line in the middle panel shows the median relation at $z=6$ for reference.} At fixed stellar mass, the distribution of magnitudes is asymmetric, with a broader spread at the bright end, and the median magnitude becomes more negative (galaxies being brighter) at higher redshifts. The scatter gets smaller from rest-frame UV to longer wavelength. {\em Bottom:} The stellar mass--magnitude relation for the same bands and redshifts. At fixed magnitude, the distribution of stellar mass is skewed toward low-mass galaxies, simply due to the fact that low-mass galaxies are more abundant in the Universe. The median stellar mass decreases by about 1\,dex from $z=12$ to 5.}
\label{fig:mag}
\end{figure*} %

\subsection{Star formation histories}
\label{sec:sfr}
Figure \ref{fig:sfh} shows the star formation histories (top panels) and the stellar mass growth histories (bottom panels) for four example galaxies in the $z=5$ halo mass range $\Mhalo=10^8$--$4\times10^{11}\,\Msun$ (as labeled in each panel). The black solid lines and red dashed lines in the top panels show the star formation rates (SFRs) averaged over 10\,Myr and 100\,Myr, respectively. These are proxies for the H$\alpha$- and UV-inferred SFRs observationally \citep[e.g.][]{sparre.2017:fire.sf.burst}. All simulated galaxies show significant `bursty' star formation histories, with starbursts occurring on time-scales of 50--100\,Myr. \referee{This feature is also seen in other cosmological zoom-in simulations with comparably high resolution and detailed physics despite different numerical methods \citep[e.g.][]{kimm.cen.2014:esc.frac,ceverino.2018:first.light.sfr}.} As discussed in Section \ref{sec:smhm}, the stellar mass can grow almost instantaneously by a factor of 2 or more at the peak of a burst, while it remains nearly constant when the SFR is low. In the least massive halo ($\Mhalo=10^8\,\Msun$ at $z=5$), the feedback from a starburst at $z\sim8$ expels most of its gas. At later times, gas accretion and cooling becomes inefficient as heating from the ionizing background becomes significant for such low mass halos \citep[e.g.][]{efstathiou.1992:uvb.suppress.sf,thoul.weinberg.1996:uvb.quench,Quinn.1996:reionization.dwarf,okamoto.2008:uvb.mass.loss,cafg.2011:dm.halo.assemby,noh.mcquinn.2014:uvb.supress.sf,sawala.2016:low.mass.halos}. Star formation in these halos is thus suppressed at later times. More massive halos are able to maintain star formation until the end of our simulation at $z=5$.

In Figure \ref{fig:sfr}, we show the SFR (left panel) and specific SFR (sSFR, right panel) averaged over the past 100\,Myr as a function of stellar mass for all galaxy snapshots in the simulated sample, color coded by their redshift. Note that the sharp upper limits at a given stellar mass in both relations are due to the fact that some galaxies form essentially all of their stellar mass in a starburst during the past 100\,Myr. We also compare our results with the observed relation for $z=6$--9 galaxies from \citet{mclure.2011:highz.galaxy.sfr} (black dashed lines). Our simulations agree well with observations at $\Ms\geq10^8\,\Msun$. At lower masses (where there are no observations), the scatter is larger due to stronger burstiness in their star formation histories, as illustrated by the solid lines in the right panel. Moreover, at fixed stellar mass, galaxies at lower redshifts have lower star formation rates on average than galaxies at higher redshifts. This trend is expected because the stellar mass growth time-scale (the ratio of stellar mass to star formation rate) of galaxies at a given redshift should be comparable to the Hubble time at that redshift \referee{and has also been found in previous studies \citep[e.g.][]{behroozi.silk.2015:hiz.gal.model,wilkins.2017:bluetides.simulation}.}

We now derive the weighted average SFR as a function of halo mass (or stellar mass) and redshift at every $\Delta\log\Mhalo=0.5$ (or $\Delta\log\Ms=0.5$) and $\Delta\log(1+z)=0.04$ for the simulated sample. All halo snapshots are included in this calculation. We then fit the results using two-dimensional linear functions
\be
\label{eqn:sfrmh}
\log\SFR = \alpha\,(\log\Mhalo-10) + \gamma\,\log\left(\frac{1+z}{6}\right) + \delta
\ee
and
\be
\label{eqn:sfrms}
\log\SFR = \alpha'\,(\log\Ms-10) + \gamma'\,\log\left(\frac{1+z}{6}\right) + \delta',
\ee
and obtain the best-fit parameters $(\alpha,\,\gamma,\,\delta)=(1.58,\,2.20,\,-1.58)$ and $(\alpha',\,\gamma',\,\delta')=(1.03,\,2.01,\,1.36)$. In Figure \ref{fig:sfhz}, we show the $\SFR(\Mhalo,\,z)$ (left panel) and $\SFR(\Ms,\,z)$ (right panel) relations derived from the simulated sample. The colors represent the weighted average SFR relative to the $z=5$ best-fit relation (in logarithmic scale, as labeled at the colorbars). This eliminates the wide dynamic range shown in the left panel of Figure \ref{fig:sfr} and allows us to see the evolution more clearly. Note that the white regions show the parameter space where our simulated sample contains no galaxies. At fixed halo mass or stellar mass, the average SFR increases by $\sim0.7\,\dex$ from $z=5$--12, in broad agreement with the qualitative trend shown in Figure \ref{fig:sfr}. 

Furthermore, after $z\sim6$, the average SFRs in low-mass galaxies (halo mass below $\Mhalo\sim10^8\,\Msun$ or stellar mass below $\Ms\sim10^4\,\Msun$, the dark blue region in Figure \ref{fig:sfhz}) are significantly lower (by about 0.6\,dex) than those inferred from the fitting functions. This is because, as mentioned above, at the end of the reionization era, the ionizing background can heat the gas in low-mass halos efficiently and prevent it from cooling and forming stars. The star formation in these galaxies is thus suppressed. We note that halos of similar masses at higher redshifts ($z\gtrsim7$) or more massive halos ($\Mhalo\gtrsim10^{8.5}\,\Msun$) at any redshift are not affected and continue normal star formation.

In Figure \ref{fig:sfh}, we show one example of such low-mass galaxies (the left most panel), where star formation is suppressed at lower redshifts. There are also halos at these masses which are completely `dark' (containing no stars). The dark halo fraction is negligible for halos above $\Mhalo=10^{8.5}\,\Msun$ at any redshift, whereas at $\Mhalo\sim10^8\,\Msun$, the dark fraction increases from less than 1\% at $z=12$ to approximate 50\% at $z=5$.\footnote{Note that the increasing dark fraction at $\Mhalo\sim10^8\,\Msun$ and below $z\sim6$ indicates that the suppression of star formation in these halos is not purely due to stellar feedback but rather points to the importance of reionization.} We will show later in Sections \ref{sec:smf} and \ref{sec:lf} that this effect leaves an imprint in the stellar mass function and luminosity functions at $z\sim6$.

Our findings are broadly in line with other simulations in the literature. For example, \citet{wise.2014:reion.esc.frac} found no dark halos above $\Mhalo\sim10^8\,\Msun$ at $z>8$. \citet{sawala.2016:low.mass.halos} found that the dark fraction decreases sharply from nearly 90\% at $\Mhalo\sim10^8\,\Msun$ to 0\% at $\Mhalo\sim10^{8.5}\,\Msun$ at $z\sim10$. They also find an increasing dark fraction with decreasing redshift at a fixed halo mass. The subtle differences are likely due to different models of the ionizing background adopted in these studies, as well as to different star formation and stellar feedback physics. \citet{wise.2014:reion.esc.frac} modeled the ionizing fields more self-consistently using radiative-hydrodynamic methods, while \citet{sawala.2016:low.mass.halos} adopted the uniform \citet{haardt.madau.2001:uvb} ionizing background at these redshifts. In addition, a dark halo only means that the expected stellar mass is lower than the mass of a few star particles. This further complicates the comparison between these results obtained at different resolutions. The effects of the ionizing radiation fields prior to complete reionization on low-mass galaxies merits future investigation.

\begin{table}
\centering
\caption{Best-fit parameters for the magnitude--stellar mass relation and the stellar mass--magnitude relation (see Section \ref{sec:mag} for details).}
\begin{threeparttable}
\begin{tabular}{ccccc}
\hline\hline
\multicolumn{5}{c}{Magnitude--stellar mass relation (Equation \ref{eqn:mstomag})} \\
\hline
\multicolumn{2}{c}{Band} & $a$ & $c$ & $d$ \\
\hline
1500\,\AA & median & -2.81 & -5.61 & -22.38 \\
1500\,\AA & 1$\sigma$ lower & -2.61 & -6.83 & -23.06 \\
1500\,\AA & 1$\sigma$ upper & -2.74 & -3.87 & -21.42 \\
B band & median & -2.63 & -3.36 & -22.46 \\
B band & 1$\sigma$ lower & -2.59 & -5.17 & -22.89 \\
B band & 1$\sigma$ upper & -2.64 & -2.52 & -22.05 \\
J band & median & -2.61 & -2.63 & -22.69 \\
J band & 1$\sigma$ lower & -2.61 & -3.86 & -23.20 \\
J band & 1$\sigma$ upper & -2.63 & -2.15 & -22.31 \\
\hline\hline
\multicolumn{5}{c}{Stellar mass--magnitude relation (Equation \ref{eqn:magtoms})} \\
\hline
\multicolumn{2}{c}{Band} & $a'$ & $c'$ & $d'$ \\
\hline
1500\,\AA & median & -0.39 & -2.59 & 8.77 \\
1500\,\AA & 1$\sigma$ lower & -0.42 & -1.65 & 8.38 \\
1500\,\AA & 1$\sigma$ upper & -0.36 & -2.99 & 9.16 \\
B band & median & -0.38 & -2.17 & 8.95 \\
B band & 1$\sigma$ lower & -0.41 & -1.59 & 8.66 \\
B band & 1$\sigma$ upper & -0.37 & -1.40 & 9.09 \\
J band & median & -0.38 & -1.85 & 8.90 \\
J band & 1$\sigma$ lower & -0.41 & -1.61 & 8.67 \\
J band & 1$\sigma$ upper & -0.38 & -0.90 & 9.01 \\
\hline\hline
\end{tabular}
\begin{tablenotes}
\item Note: All magnitudes are derived from intrinsic stellar luminosities without accounting for dust attenuation and nebular line emission.
\end{tablenotes}
\end{threeparttable}
\label{tbl:magfit}
\end{table}

\subsection{Broad-band photometry}
\label{sec:mag}
We use the BPASSv2.0 stellar population synthesis models to calculate the broad-band luminosities and magnitudes for the simulated galaxies, using the binary models with a \citet{kroupa.2002:imf} IMF from 0.1--$100\,\Msun$ as our default model. We only consider intrinsic stellar continuum here, and ignore dust extinction and strong nebular line emission in the rest-frame UV and optical, as well as dust re-emission in the infrared (IR). We will explore the effect of dust attenuation in Section \ref{sec:extinction}. In Figure \ref{fig:mag}, we show the magnitude--stellar mass relation (top panels) for our simulated sample with $\Ms>10^{3.5}\,\Msun$ for three example redshift and band combinations. We also show the inverse relation, the stellar mass--magnitude relations for the same combinations in the bottom panels. Only galaxies brighter than ${\rm M_{AB}}=-10$ are shown to ensure that our simulations are complete. The two-dimensional histogram represents the total {\em weight} (as defined in Section \ref{sec:weights}) of all galaxy snapshots in a pixel in logarithmic scale. This reflects the `correct' relative number of galaxies in the Universe between pixels\footnote{We remind the reader that we include all snapshots in the analysis to account for time variability of galaxy properties, which is important for UV luminosities, but we may underestimate the scatter for halos above $\Mhalo>10^{11}\,\Msun$ where our sample does not contain large number of independent halos (see Section \ref{sec:definition} for details).}. The red solid and dashed lines illustrate the best-fit median relation and $1\sigma$-scatter (16--84 per cent) as obtained below.

At fixed stellar mass, the distribution of magnitudes in a specific band tends to be asymmetric, with a broader spread at the bright end. The asymmetry is driven by the evolution of stellar populations: the luminosity of a stellar population declines rapidly as the most massive stars die (in about 3--30\,Myr). Therefore, the luminosity of a galaxy depends not only on its total stellar mass but also on its recent star formation history. This feature is more prominent in low-mass galaxies which have significant bursty star formation histories. Figure \ref{fig:mag} also shows that this effect is strongest in the rest-frame UV where young stars overwhelmingly dominate the starlight and becomes weaker at longer wavelengths, as rest-frame optical B-band relation has smaller scatter than that of rest-frame 1500\,\AA.

Furthermore, galaxies at higher redshifts appear brighter on average than those of similar stellar masses at lower redshifts, simply due to the fact that high-redshift galaxies have younger stellar populations and higher ongoing SFRs. We parametrize the magnitude--stellar mass relation with a linear function
\be
\label{eqn:mstomag}
{\rm M_{AB,\,band}} = a\,(\log\Ms-10) + c\,\log\left( \frac{1+z}{6} \right) + d,
\ee
where we assume a fixed slope $a$ at any redshift but a redshift-dependent normalization to capture this feature. We fit the weighed median, $1\sigma$ lower- and upper-bound relations (above $\Ms=10^{3.5}\,\Msun$) obtained from eight subsamples in different redshift intervals from $z=5$--12 all together to determine the parameters for a given band. The top panels of Figure \ref{fig:mag} illustrate three examples of this relationship, and we list the best-fit parameters for rest-frame 1500\,\AA, B band, and J band in the top half of Table \ref{tbl:magfit}.

We similarly assume a linear function for the stellar mass--magnitude relation
\be
\label{eqn:magtoms}
\log\Ms = a' \,({\rm M_{AB,\,band}}+20) + c' \,\log\left( \frac{1+z}{6} \right) + d' ,
\ee
and fit the weighted median and $1\sigma$ relations for galaxies brighter than ${\rm M_{AB,\,band}}=-10$ to obtain the parameters. Some examples are shown in Figure \ref{fig:mag} and the best-fit parameters for rest-frame 1500\,\AA, B band, and J band are listed in Table \ref{tbl:magfit} (the bottom block). We emphasize that the two relations are fundamentally different from each other -- the distribution of stellar mass at fixed magnitude is biased toward low-mass galaxies, simply because they have much higher number densities in the Universe than more massive galaxies (see also Section \ref{sec:smf}). The stellar mass--magnitude relation is also redshift-dependent, with the median stellar mass decreasing by about 1\,dex from $z=12$ to 5 at a given magnitude.

\begin{figure}
\centering
\includegraphics[width=\linewidth]{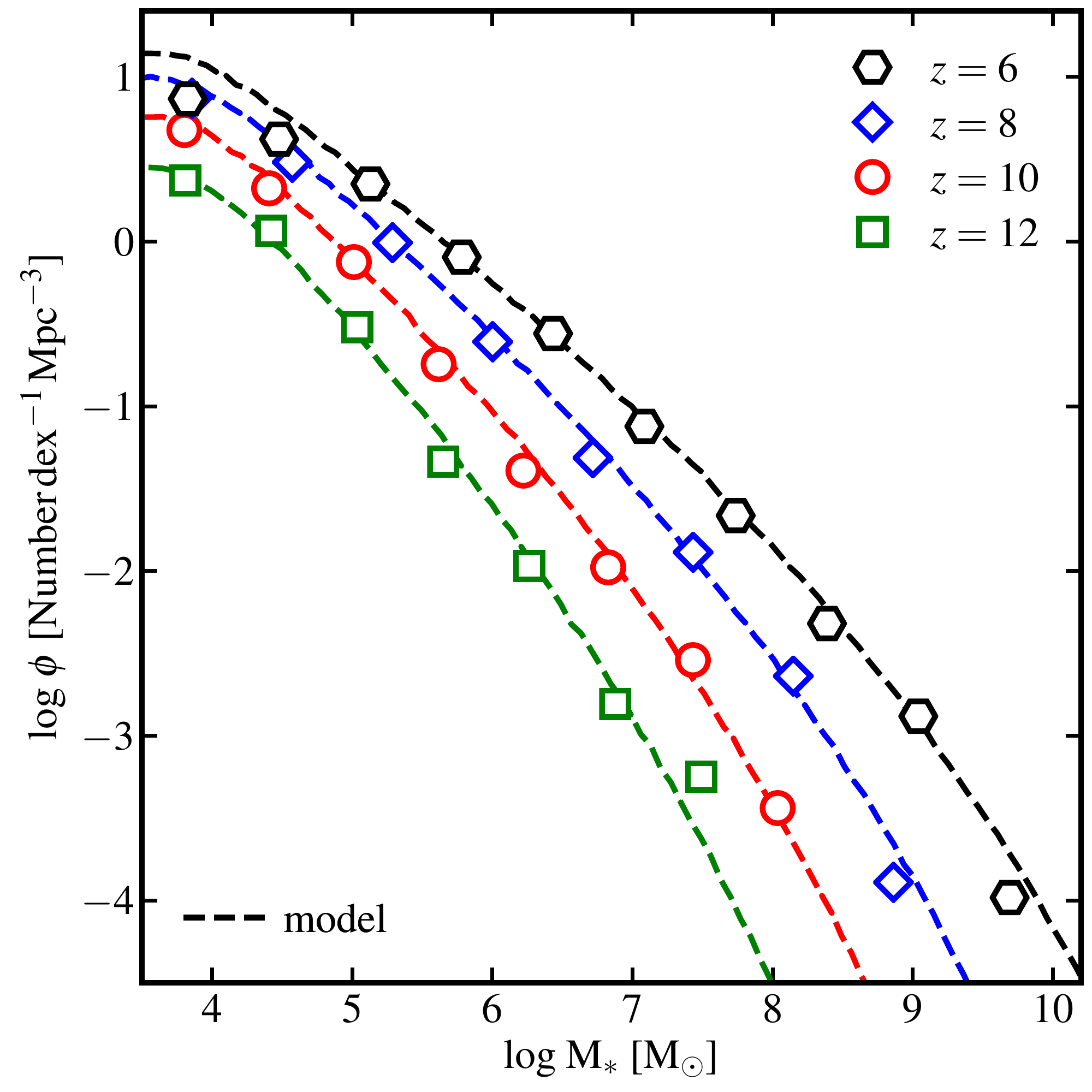} 
\caption{Predicted stellar mass functions above $\Ms=10^{3.5}\,\Msun$ at $z=6$, 8, 10, and 12. The open symbols show the results derived from the simulated sample using the weights constructed in Section \ref{sec:weights}. The dashed lines show the model stellar mass functions from convolving between the stellar mass--halo mass relation from Section \ref{sec:smhm} and the halo mass function, assuming each halo contains one central galaxy. Both methods only account for halos more massive than $\Mhalo=10^{7.5}\,\Msun$. The two stellar mass functions agree well with each other for a broad range of mass and redshift. The low-mass-end slope steepens with increasing redshift (from $\alpha=-1.83$ at $z\sim6$ to $\alpha=-2.18$ at $z\sim12$). At $z\sim6$, the stellar mass function derived from simulations flattens and falls below the model stellar mass function by a factor of 2 below $\Ms\sim10^{4.5}\,\Msun$, owing to the 50\% fraction of dark halos around $\Mhalo\sim10^8\,\Msun$. A comparison with observations is shown later in Section \ref{sec:comp} (Figure \ref{fig:comp}). We make our predictions publicly available (see Appendix \ref{app:smf} for details).}
\label{fig:smf}
\end{figure} %

\begin{figure*}
\centering
\includegraphics[width=\linewidth]{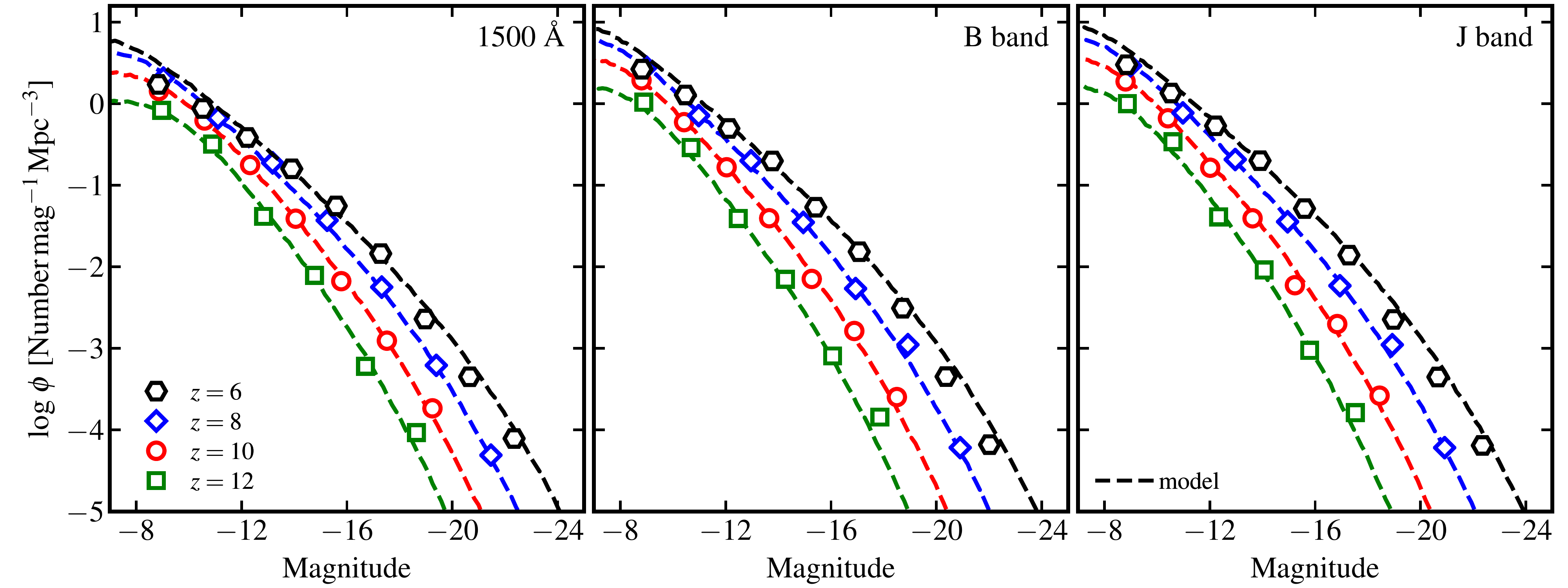} 
\caption{Predicted luminosity functions at rest-frame 1500\,{\AA} (left), B band (middle), and J band (right) at $z=6$, 8, 10, and 12. All magnitudes are intrinsic without accounting for dust attenuation. The open symbols show the results derived from the simulated catalog. The dashed lines show the model luminosity functions from convolution between the magnitude--stellar mass relation in Section \ref{sec:mag} and the model stellar mass functions in Section \ref{sec:smf}. As in Figure \ref{fig:smf}, the faint-end slope steepens with increasing redshift. The stronger flattening of the $z=6$ luminosity functions fainter than $\rm M_{AB}\sim-12$ is due to the suppression of star formation in halos around $\Mhalo\sim10^8\,\Msun$ by the strong ionizing background. We will compare our predicted UV luminosity function with observations later in Section \ref{sec:comp} and Figure \ref{fig:comp}. We make our predictions publicly available (see Appendix \ref{app:smf} for details).}
\label{fig:lf}
\end{figure*} %

\section{Stellar mass functions in the early universe}
\label{sec:smf}
Now we calculate the stellar mass function using two distinct approaches. First, we utilize the  weights constructed in Section \ref{sec:weights}: at a certain redshift, we collect all halo snapshots within $\Delta z=\pm0.5$ from our simulated catalog. We then add the weights of galaxies in stellar mass bins and divide $\sum_i w_i$ by the comoving volume corresponding to the $\Delta z=\pm0.5$ redshift interval. Only halos above $\Mhalo=10^{7.5}\,\Msun$ are taken into account. In Figure \ref{fig:smf}, we show the results in number\,$\rm dex^{-1}\,Mpc^{-3}$ above $\Ms=10^{3.5}\,\Msun$ at $z=6$, 8, 10, and 12 with the open symbols. The data are tabulated in Tables \ref{tbl:smf} and \ref{tbl:lf} for readers to use. 

Alternatively, we can model the stellar mass function by directly convolving the stellar mass--halo mass relation derived in Section \ref{sec:smhm} with the halo mass function at a given redshift. We use a Monte Carlo method: {we generate a large number of mock halos more massive than $\Mhalo=10^{7.5}\,\Msun$ with number densities following the halo mass function, assign each halo a stellar mass as described below, and derive the stellar mass function from the mock catalog}. We assume a) every halo hosts one galaxy (considering only central galaxies) and b) the stellar mass follows a lognormal distribution at a given halo mass, with the median and $1\sigma$ dispersion following Equations \ref{eqn:smhm} and \ref{eqn:scatter}. In this calculation, we use Equations \ref{eqn:smhm} and \ref{eqn:scatter} at all redshifts and all halo masses, but we caution that uncertainties may arise at the high-mass end (see discussion in Section \ref{sec:smhm}). The results are shown by dashed lines in Figure \ref{fig:smf}. For simplicity, we refer them as model stellar mass functions thereafter. We also make these results publicly available (see Appendix \ref{app:smf} for details). We will compare our predictions with observations later in Section \ref{sec:comp}.

We highlight the following features shown in Figure \ref{fig:smf}. (1) The low-mass end of the stellar mass function (asymptotic form $\phi\,\dd\log\Ms\sim\Ms^{\alpha+1}\,\dd\log\Ms$) steepens with increasing redshift, with the slope decreasing from {$\alpha=-1.80\pm0.02$} at $z=6$ to {$\alpha=-2.13\pm0.12$} at $z=12$. The evolution of the slope is robust, although the exact slope at a given redshift may vary according to how it is computed\footnote{The slopes quoted here are obtained by fitting the stellar mass functions derived from the simulated catalog with a \citet{schechter.1976:schechter.funciton} function. We also experiment with fitting the model mass functions in different dynamic ranges or using a double power-law function. The slope obtained at a given redshift varies systematically with method by about 0.2.}. This trend is consistent with the observed stellar mass functions \citep[e.g.][]{song.2016:stellar.mass.func.z4to8} \referee{and has been widely reproduced in cosmological simulations \citep[e.g.][]{wilkins.2016:bluetides.photometry,ceverino.2017:firstlight.project}}. Such a feature is directly inherited from the halo mass functions, which also steepen with increasing redshift at the low-mass end \citep[e.g.][]{reed.2003:evolution.hmf}. (2) The model stellar mass functions agree well with those derived from the simulated catalog for a broad range of stellar mass and redshift. This demonstrates that the fitting functions in Section \ref{sec:smhm} describe the stellar mass--halo mass relation for the simulated sample reasonably well. (3) The discrepancies between the two stellar mass functions in the highest-mass bin is due to small numbers of galaxies in the simulated sample at the high-mass end. (4) The apparent flattening of the model stellar mass functions at $\Ms\sim10^{3.5}\,\Msun$ is an expected artifact because we exclude all halos below $\Mhalo=10^{7.5}\,\Msun$. 

More importantly, the $z=6$ stellar mass function derived from the simulated sample shows a flattening below $\Ms\sim10^{4.5}\,\Msun$ and falls below the model mass function by a factor of 2. This is caused by the 50\% fraction of dark halos at $\Mhalo\sim10^8\,\Msun$ at $z\sim6$. In other words, the assumption we adopted in the model that every halo hosts one galaxy breaks down at $\Mhalo\sim10^8\,\Msun$. Note that if we ignore all the `dark halos' in the simulated catalog and repeat the exercise, the two $z=6$ stellar mass functions agree well with each other. The large dark fraction in low-mass halos at lower redshifts is because of the suppression of star formation by the ionizing background near the end of reionization (see Section \ref{sec:sfr}). The stellar mass functions at higher redshift do not show such flattening. This effect may relieve the tension between the number of low-mass galaxies in the Local Group and that needed for cosmic reionization \citep[e.g.][]{bullock.2000:reionization.satellites,somerville.2002:squelching,mbk.2014:reion.faint.galaxy}.

\begin{figure}
\centering
\includegraphics[width=\linewidth]{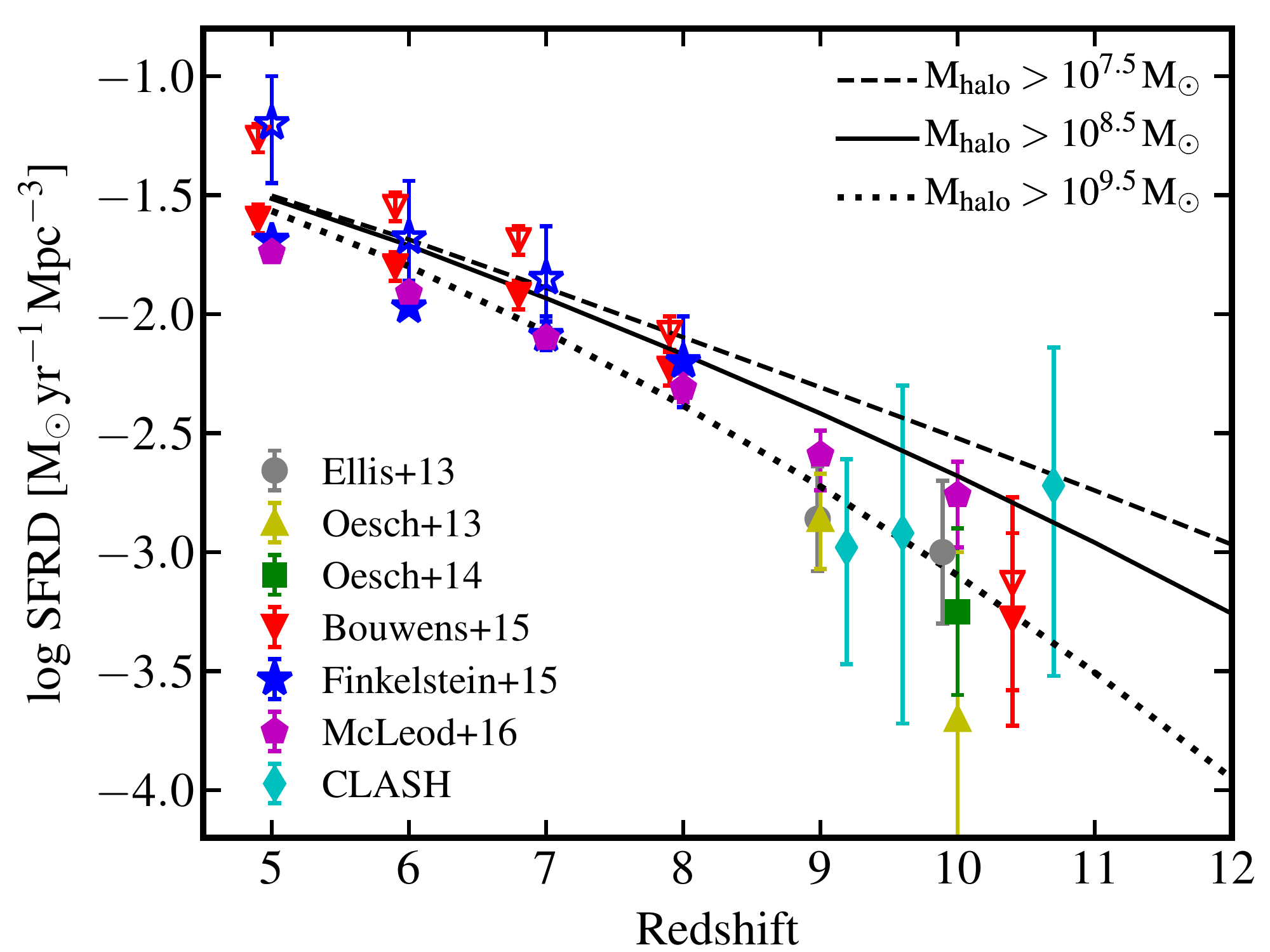} 
\caption{Cosmic star formation rate density (SFRD) from $z=5$--12. The lines are derived by convolving the SFR--$\Mhalo$ relation given by Equation \ref{eqn:sfrmh} and the halo mass function at the same redshift. The dashed, solid, and dotted lines show the results obtained by integrating over the halo mass range as labeled. Observationally inferred SFRDs from the literature are shown with symbols and errorbars. These results are derived by integrating the best-fit UV luminosity functions brighter than $\rm M_{UV}\sim-17$. Data corrected (uncorrected) for dust attenuation are shown by open (filled) symbols. At $z\lesssim8$, our predictions agree with data within observational uncertainties. The $z\gtrsim9$ SFRD is poorly constrained due to the small size of each observational sample. Our simulations suggest that low-mass halos dominate the SFRD, due to their rapidly increasing number densities at these redshifts. This is beyond the detection limits of current observational facilities ($\rm M_{UV}\sim-17$, roughly corresponding to halo mass $\Mhalo\sim10^{9.5}\,\Msun$ in our simulations). Future deep surveys by JWST will be able to put stronger constrains on the $z\gtrsim9$ SFRD.}
\label{fig:sfrd}
\end{figure} %

\section{Multi-band luminosity functions}
\label{sec:lf}
We calculate the luminosity functions at several bands following the same method described in Section \ref{sec:smf}. Again, only halos above $\Mhalo=10^{7.5}\,\Msun$ are taken into account. In Figure \ref{fig:lf}, we show the results in number\,$\rm mag^{-1}\,Mpc^{-3}$ at $z=6$, 8, 10, and 12 (the open symbols). The data are also provided in Appendix \ref{app:smf}. We also model the luminosity functions using a Monte Carlo method by convolving the magnitude--stellar mass relation derived in Section \ref{sec:mag} (Equation \ref{eqn:mstomag}) with the model stellar mass functions in Section \ref{sec:smf} in a similar way described above. The model luminosity functions are shown by the dashed lines in Figure \ref{fig:lf}. Note that we only consider intrinsic stellar continuum emission here, but we will explore the effect of dust extinction in Section \ref{sec:extinction} and compare with observations in Section \ref{sec:comp}. We do not model nebular line emission in this paper.

The luminosity functions derived from the simulations agree well with models for a broad range of magnitude and redshift. Again, the faint-end slope (asymptotic form $\phi\,\dd {\rm M_{AB,\,band}} \sim 10^{-0.4(\alpha+1){\rm M_{AB,\,band}}} \dd {\rm M_{AB,\,band}}$) steepens with increasing redshift (e.g. from {$\alpha=-1.85\pm0.06$} at $z=6$ to {$\alpha=-2.17\pm0.10$} at $z=12$ at 1500\,\AA). The trend is in good agreement with observations \citep[e.g.][]{bouwens.2015:uvlf.z4to10,finkelstein.2015:uvlf.combined.field}, \referee{semi-analytic galaxy formation models \citep[e.g.][]{clay.2015:sam.high.redshift.uvlf,cowley.2018:sam.jwst.deep.survey}, and other simulations \citep[e.g.][]{gnedin.2016:croc.faint.end.uvlf,ceverino.2017:firstlight.project,wilkins.2017:bluetides.simulation} at these redshifts, but the exact slopes depend largely on the fitting method}. The flattening at the faintest bin at any redshift is due to the incompleteness of halos below $\Ms=10^{7.5}\,\Msun$. Similarly, the luminosity functions show a flattening below $\rm M_{1500\,\text{\AA}}\sim-12$, $\rm M_{B}\sim-12$, and $\rm M_{J}\sim-12$ at $z=6$, {as seen from the fact that the luminosity functions derived from the simulated catalog fall below the model luminosity functions roughly by a factor of 2}. This is caused by the large fraction of dark halos and the rapid drop in SFR at $z\sim6$ below halo mass $\Mhalo\sim10^8\,\Msun$, where star formation is suppressed by the ionizing background (Figure \ref{fig:sfhz}).

\begin{figure}
\centering
\includegraphics[width=\linewidth]{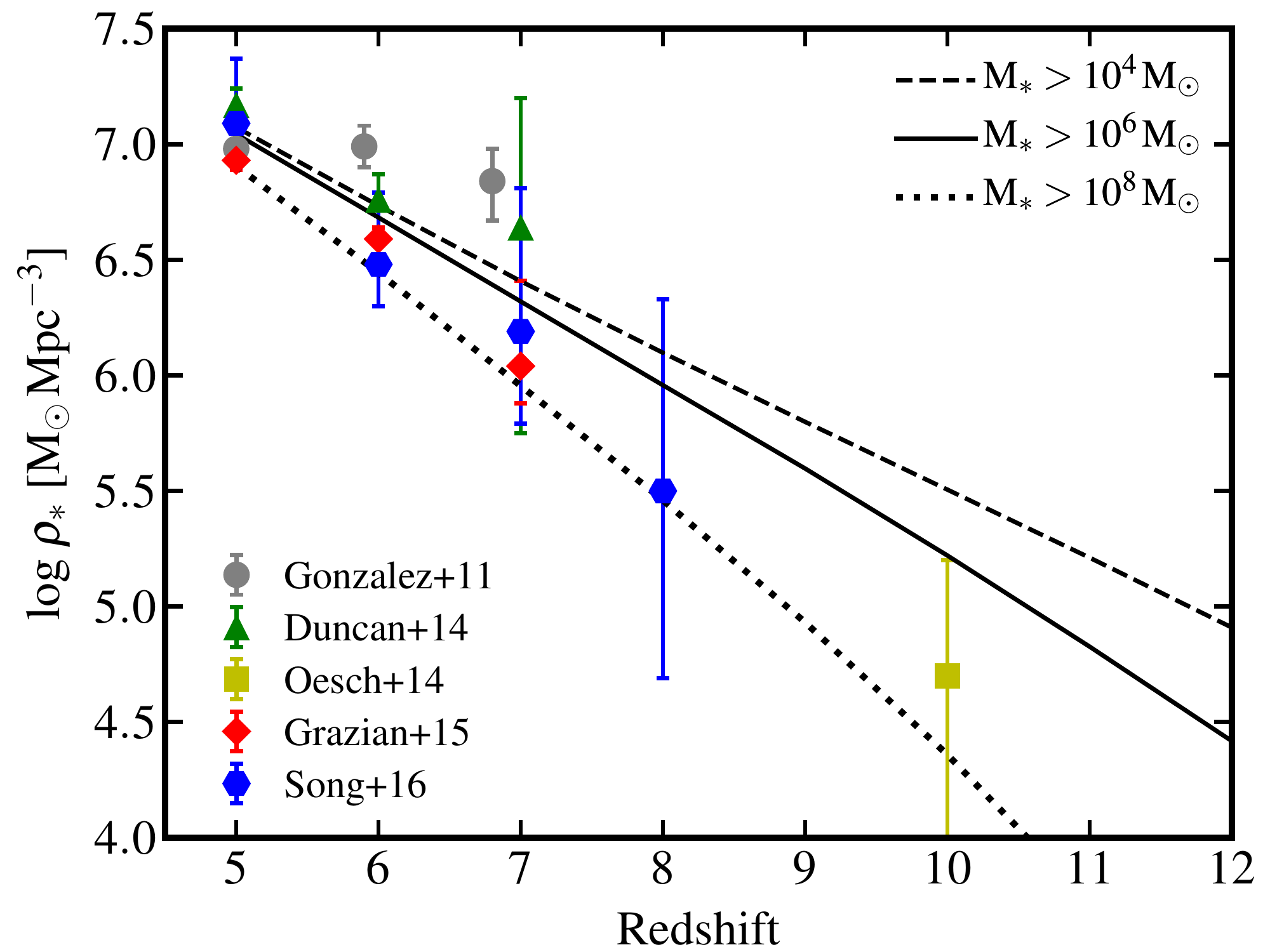} 
\caption{Stellar mass density from $z=5$--12. The lines are derived by integrating the model stellar mass functions in Section \ref{sec:smf} over the mass range as labeled. At high redshift, the stellar mass density is dominated by low-mass galaxies, due to the rapid steepening of the low-mass-end of the stellar mass function. Observationally inferred data from the literature are shown by filled symbols with errorbars. These observational results are derived by integrating the best-fit stellar mass functions above $\Ms>10^8\,\Msun$. Our equivalent prediction (the dotted line) broadly agrees well with more recent studies. The discrepancies between these measurements likely originate from systematic uncertainties in stellar mass measurements.}
\label{fig:smd}
\end{figure} %

\section{Discussion}
\label{sec:discussion}

\subsection{SFR and stellar mass densities at $z\geq5$}
\label{sec:smd}
We now derive the cosmic SFR density (SFRD) from $z=5$--12 by convolving the SFR--$\Mhalo$ relation given by Equation \ref{eqn:sfrmh} (the average SFR at a given halo mass) and the halo mass function at the same redshift. In Figure \ref{fig:sfrd}, we show the results obtained by integrating over the halo mass range above $\Mhalo=10^{7.5}\,\Msun$ (dashed), $\Mhalo=10^{8.5}\,\Msun$ (solid), and $\Mhalo=10^{9.5}\,\Msun$ (dotted) to $\Mhalo=10^{12}\,\Msun$. The contributions from more massive halos are negligible at these redshifts (because they are extremely rare). In Figure \ref{fig:sfrd}, we also show observationally inferred SFRDs from \citet{ellis.2013:hudf12}, \citet{oesch.2013:galaxies.z9to12}, \citet{oesch.2014:uvlf.sfrd.zgt9}, \citet{bouwens.2015:uvlf.z4to10}, \citet{finkelstein.2015:uvlf.combined.field}, \citet{mcloed.2016:z9.z10.galaxy.hff.clash}, and CLASH detections \citep{zheng.2012:clash.z10.galaxy,coe.2013:clash.candidate.z11,bouwens.2014:clash.z9.candidate}. Data corrected (uncorrected) for dust attenuation are shown by open (filled) symbols\footnote{{We note that the conversion between rest-frame UV luminosity and SFR and the amount of dust correction are still very uncertain.}}. At $z\lesssim8$, our predictions broadly agree with data within observational uncertainties. The SFRD at $z\gtrsim9$ are still poorly constrained observationally. These results are derived by integrating the best-fit UV luminosity functions brighter than $\rm M_{UV}\sim-17$. This limit does not correspond to a unique halo mass, but is roughly consistent with what we obtain by integrating down to $\Mhalo=10^{9.5}\,\Msun$ at these redshifts (cf. Figures \ref{fig:smhm} and \ref{fig:mag}). Note that the number of galaxies in the observed $z\gtrsim9$ sample is small, and some works are based on single galaxy detections. Our simulations suggest that the majority of star formation takes place in halos below $\Mhalo=10^{9.5}\,\Msun$ at $z\gtrsim9$, but these low-mass galaxies are too faint to be detectable with current observational facilities. This may account for the apparent rapid decline in SFRD at these redshifts \citep[e.g.][]{oesch.2014:uvlf.sfrd.zgt9}. Future deep surveys by JWST at these redshifts are expected to put strong constraints on the $z\geq9$ SFRD. 

We also calculate the stellar mass density from $z=5$--12 by integrating the model stellar mass functions in Section \ref{sec:smf} in certain stellar mass intervals. The three lines in Figure \ref{fig:smd} show the results for $\Ms>10^6\,\Msun$ (solid), $\Ms>10^4\,\Msun$ (dashed), and $\Ms>10^8\,\Msun$ (dotted). At these redshifts, high-mass galaxies ($\Ms>10^{10}\,\Msun$) only contribute a negligible fraction (less than 0.05\,dex) of the total stellar mass due to their low number densities, so the total stellar mass density is insensitive to our uncertainties in the high-mass end of the stellar mass functions. In Figure \ref{fig:smd}, we also compare our predictions with observationally inferred results in the literature \citep[symbols with errorbars, including][]{gonzalez.2011:smf.z4to7,duncan.2014:evolution.smf.z4to7,oesch.2014:uvlf.sfrd.zgt9,grazian.2015:stellar.mass.function,song.2016:stellar.mass.func.z4to8}. Note that these results are derived by integrating the best-fit stellar mass functions above $\Ms>10^8\,\Msun$. Our predictions (the dotted line, which uses the same limit) broadly agree with more recent studies \citep{oesch.2014:uvlf.sfrd.zgt9,grazian.2015:stellar.mass.function,song.2016:stellar.mass.func.z4to8}. We note that although some groups report consistent stellar mass densities at these redshifts, their stellar mass functions do not usually agree with each other (see Section \ref{sec:comp} or figure 9 in \citealt{song.2016:stellar.mass.func.z4to8}). 

\subsection{Dust extinction in rest-frame UV}
\label{sec:extinction}
\begin{figure}
\centering
\includegraphics[width=\linewidth]{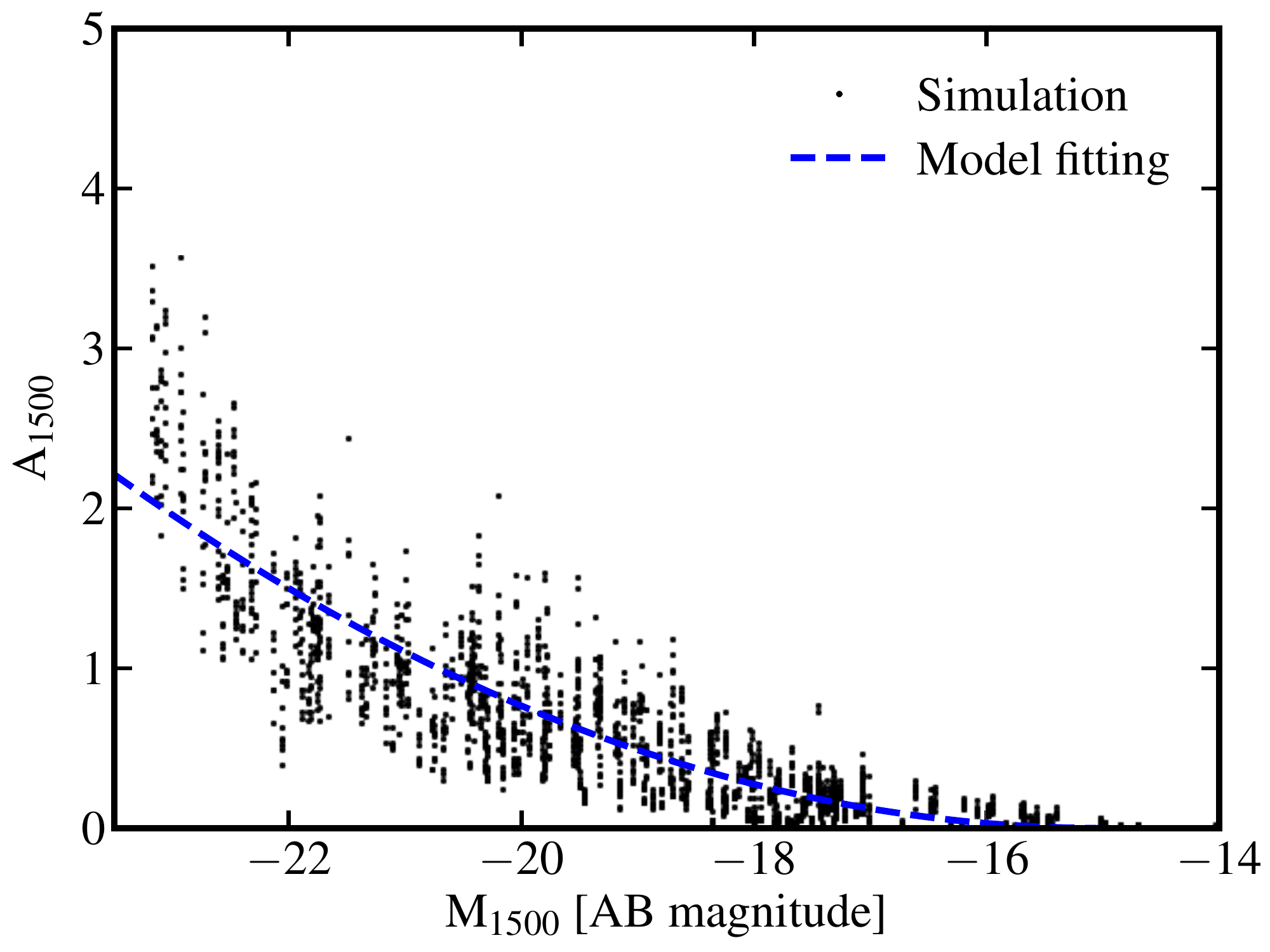} 
\caption{The relation between UV extinction and intrinsic UV magnitude in the simulations. The points show all simulated galaxies from $z=5$--8, with ten different sightlines for each galaxy. We determine the dust extinction by ray-tracing and assuming a constant dust-to-metal ratio, using the distribution of gas and metals in the simulations. The blue dashed line shows the best-fit relation in Equation \ref{eqn:extinction}.}
\label{fig:extinction}
\end{figure}

So far we have only focused on intrinsic luminosity of our simulated galaxies, while dust obscuration can be very important in relatively massive galaxies \citep[e.g.][]{cen.kimm:z7.galaxy.ir.properties,cullen.2017:dust.attenuation.z5,wilkins.2017:bluetides.simulation}. In this section, we estimate the amount of dust extinction in the rest-frame UV band. We follow the method from \citet{hopkins.2005:quasar.lifetime} and calculate the extinction by ray-tracing the emission from star particles including dust attenuation self-consistently from the dust and metals in the simulation \citep[see also][]{hopkins.2014:fire.galaxy,feldmann.2016:fire.quenching.letter,feldmann.2017:massive.fire.long}. We assume a canonical dust-to-metal ratio of 0.4 \citep{dwek.1998:dust.abundance}. Following a Small Magellanic Cloud-like extinction curve from \citet{pei.1992:reddening.curve}, we obtain a dust opacity of $2.06\times10^3\,{\rm cm^2\,g^{-1}}$ at 1500\,{\AA} at solar metallicity. For each simulated galaxy, we include all gas particles within $\frac{1}{2}\Rmax$ (about 1.5 times the size of the stellar component, see Section \ref{sec:definition}) and calculate the extinction by ray-tracing from every star particle to a hypothetical observer along ten different sightlines. Note that the gas in these high-redshift galaxies is clumpy, so the extinction can differ by several magnitudes between sightlines. In Figure \ref{fig:extinction}, we show the relation between extinction A$_{1500}=-2.5\log(F_{1500}/F_{1500,\,0})$ and intrinsic UV magnitude M$_{1500}$ for all sightlines and all simulated galaxies from $z=5$--8. We do not find significant redshift dependence in our sample, but we caution that this may be due to the small sample size at the massive end. We fit the results with a parabolic function
\be
\label{eqn:extinction}
{\rm A_{1500}} = (0.0306{\pm0.0002})\,({\rm M_{1500}}+15)^2,
\ee
and quote a uniformly distributed scatter with half-width $\Delta{\rm A_{1500}}=-0.125\,({\rm M_{1500}}+15)$ at ${\rm M_{1500}}<-15$. A full radiative transfer calculation of dust extinction, scatter, and re-emission is beyond the scope of the current paper, but will be the subject of a future study.

\begin{figure}
\centering
\includegraphics[width=\linewidth]{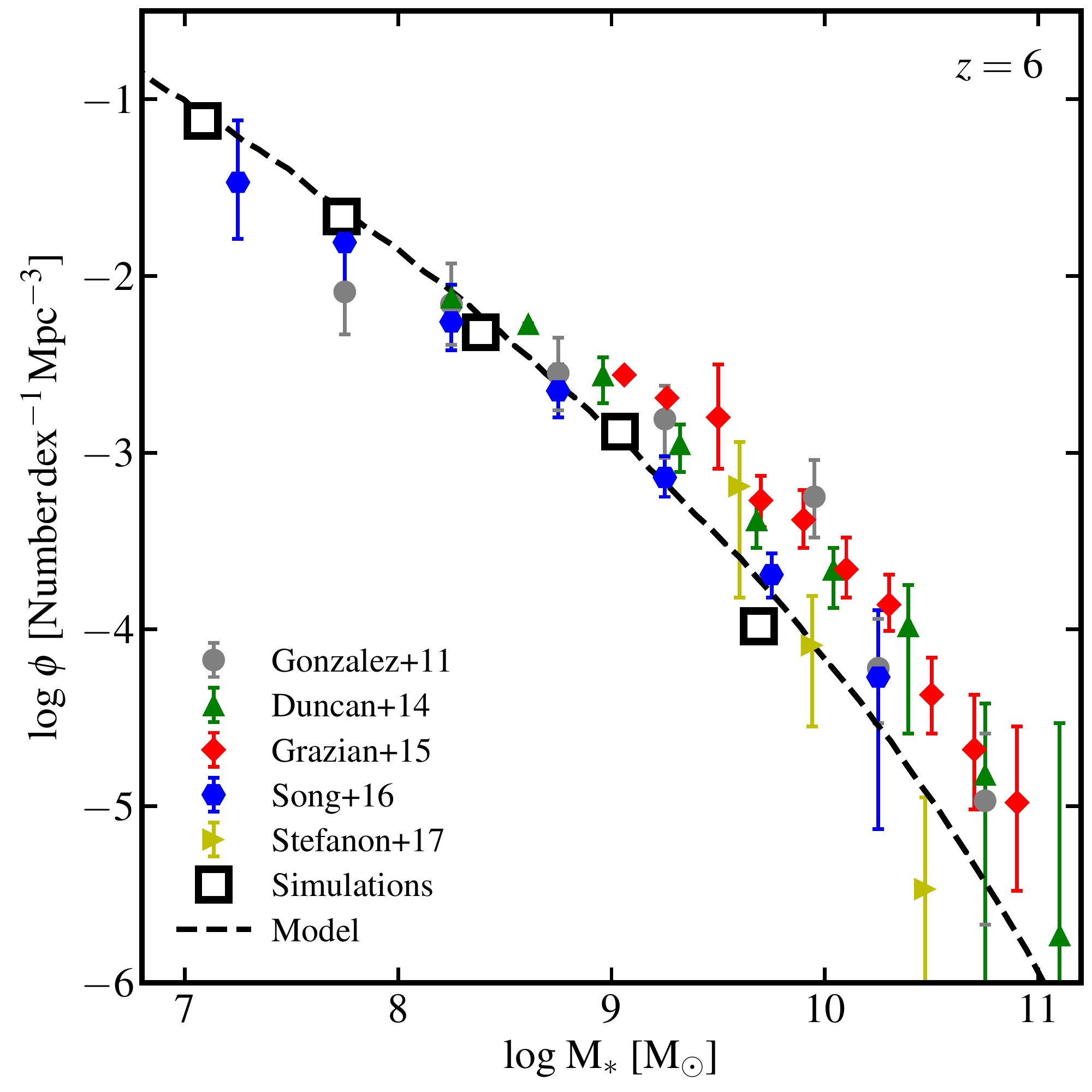}
 \includegraphics[width=\linewidth]{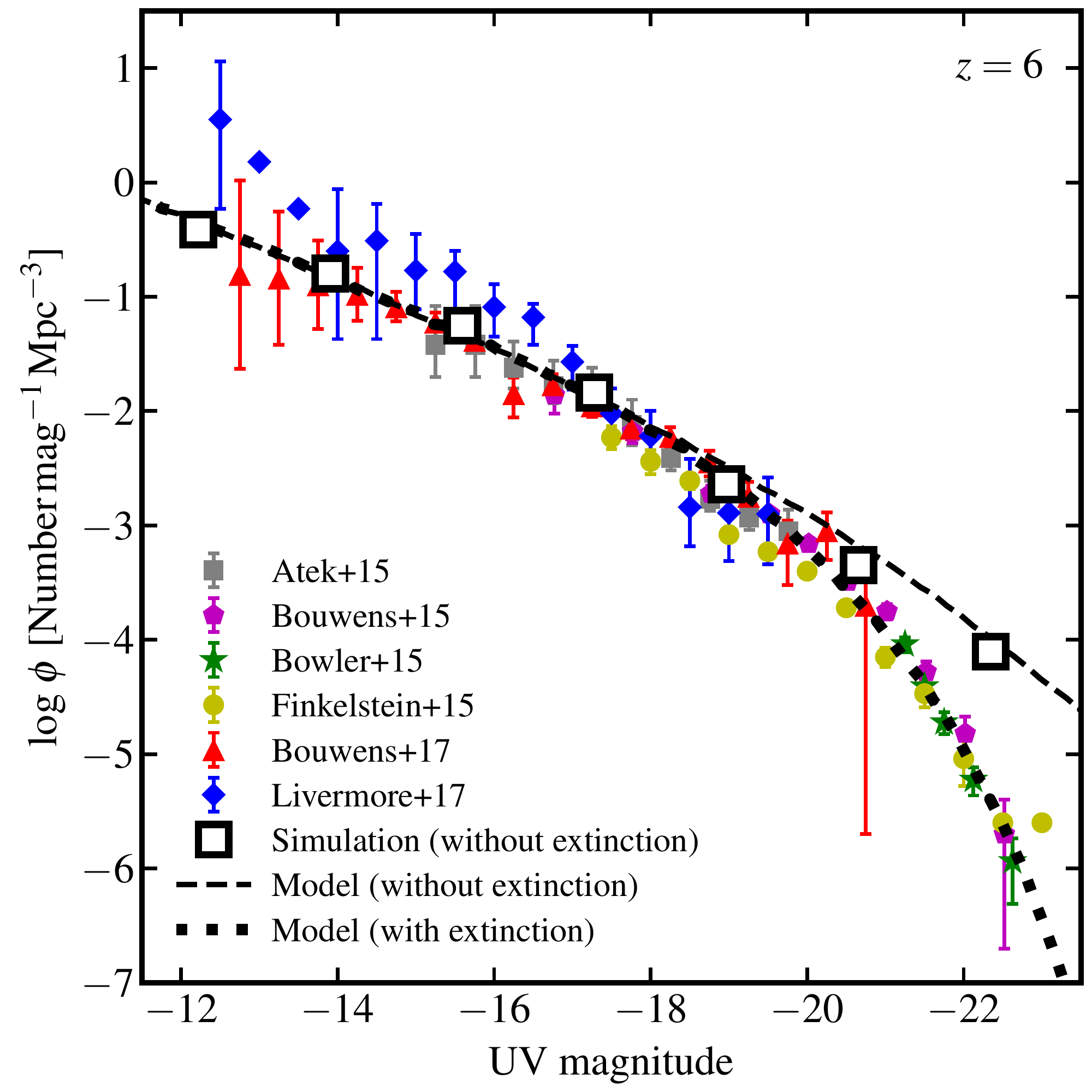}
\caption{{\em Top:} The stellar mass function at $z=6$. The open squares and the dashed line show the simulation-derived and the model stellar mass functions obtained in Section \ref{sec:smf} (the same as in Figure \ref{fig:smf}). Symbols with errorbars show a compilation of observations from \citet{gonzalez.2011:smf.z4to7}, \citet{duncan.2014:evolution.smf.z4to7}, \citet{grazian.2015:stellar.mass.function}, \citet{song.2016:stellar.mass.func.z4to8}, and \citet{stefanon.2017:hiz.optical.lf}. Our results are broadly consistent with observations. The discrepancies might be due to systematic uncertainties in deriving stellar mass from single-band magnitude, incompleteness corrections at the low-mass end, and cosmic variance at the massive end. {\em Bottom:} The UV luminosity at $z=6$. A compilation of observations from \citet{atek.2015:uvlf.z6to8.combined}, \citet{bowler.2015:galaxy.uvlf.z6}, \citet{bouwens.2015:uvlf.z4to10}, \citet{finkelstein.2015:uvlf.combined.field}, \citet{bouwens.2017:lensing.uncertainty}, and \citet{livermore.2017:faint.galaxies} are shown by symbols with errorbars. Using intrinsic luminosities, the model tends to predict higher number densities than observed at the bright end. The thick dashed line shows the luminosity function after accounting for the dust extinction in the simulations (see text for details). The good agreement with data suggests that the turnover at the bright-end of the UV luminosity function is largely due to dust extinction. Approximately 37\% (54\%) of the UV luminosity from galaxies brighter than $\rm M_{1500}=-13$ ($-17$) is obscured by dust at $z\sim6$.}
\label{fig:comp}
\end{figure} %

\subsection{Comparison with observations}
\label{sec:comp}
In this section, we compare our predicted stellar mass functions and luminosity functions with observations. In the top panel in Figure \ref{fig:comp}, we show the $z=6$ stellar mass function derived from the simulated catalog (open squares) and from direct convolution between the stellar mass--halo mass relation and the halo mass function (dashed lines). They are identical to those in Figure \ref{fig:smf}, but we only show $\Ms\geq10^7\,\Msun$ where the observational results are available. We also show a compilation of observations from \citet{gonzalez.2011:smf.z4to7}, \citet{duncan.2014:evolution.smf.z4to7}, \citet{grazian.2015:stellar.mass.function}, \citet{song.2016:stellar.mass.func.z4to8}, and \citet{stefanon.2017:hiz.optical.lf} (symbols with errorbars). Above $\Ms\sim10^9\,\Msun$, our model agrees well with \citet{song.2016:stellar.mass.func.z4to8} and \citet{stefanon.2017:hiz.optical.lf}, but falls below some other results. Below $\Ms\sim10^8\,\Msun$, we predict slightly higher number densities than \citet{song.2016:stellar.mass.func.z4to8}.

In the bottom panel, we show the $z=6$ UV luminosity function (rest-frame 1500\,\AA) from our predictions (open squares and the thin dashed line, identical to those in Figure \ref{fig:lf}) and from observations in \citet{atek.2015:uvlf.z6to8.combined}, \citet{bowler.2015:galaxy.uvlf.z6}, \citet{bouwens.2015:uvlf.z4to10}, \citet{finkelstein.2015:uvlf.combined.field}, \citet{bouwens.2017:lensing.uncertainty}, and \citet{livermore.2017:faint.galaxies} (symbols with errorbars). First, we only consider the intrinsic stellar luminosities without accounting for dust extinction (the thin dashed line), which results in the fact that our model predicts higher number densities than observed at the bright end. 

To quantify the effect of  dust attenuation, we use a Monte Carlo method to apply the dust attenuation determined in Section \ref{sec:extinction} to the model UV luminosity function. We adopt the median attenuation from Equation \ref{eqn:extinction}, with a magnitude-dependent scatter following a uniform distribution with half-width $\Delta{\rm A_{1500}}=-0.09375\,({\rm M_{1500}}+15)$ at ${\rm M_{1500}}<-15$. The model UV luminosity function after dust extinction is shown by the thick dashed line in Figure \ref{fig:comp}, which agrees surprisingly well with observations at the bright end. This result suggest that the bright-end of the UV luminosity function is mostly set by dust obscuration, \referee{in line with predictions from semi-analytic models \citep{somerville.2012:galaxies.uv.to.ir} and cosmological simulations \citep{wilkins.2017:bluetides.simulation}}. We find that dust extinction becomes significant for galaxies with intrinsic UV magnitude brighter than $\rm M_{1500}\sim-20$. The star formation in these galaxies cannot be fully probed in the rest-frame UV. Approximately 37\% of the UV light from galaxies brighter than $\rm M_{1500}=-13$ at $z=6$ is obscured by dust according to our model. The obscured fraction is 54\% if only galaxies brighter than $\rm M_{1500}=-17$ are considered. These numbers are broadly in line with observational estimates of the dust obscured fraction of star formation at these redshifts \citep[see e.g.][]{finkelstein.2015:uvlf.combined.field,bouwens.2015:uvlf.z4to10}.

Our predicted UV luminosity function (after dust attenuation) is in good agreement with current data in a broad range of magnitudes\footnote{Note that the sample in \citet{livermore.2017:faint.galaxies} has only one galaxy in the faintest bin, and no galaxy in the next two bins (upper limits).}, but the predicted stellar mass function shows considerable discrepancies with observational measurements. We note that the stellar mass functions from different groups also do not in general agree perfectly with each other. We discuss several systematic uncertainties that might be important in these measurements. {First, a non-negligible fraction of the light from galaxies will be missed due to the finite surface brightness depth of an observational campaign. Therefore, the stellar mass of a galaxy is possibly underestimated. This effect becomes much stronger at lower masses \citep[e.g.][]{ma.2018:hizfire.sizes}.} Second, the incompleteness correction at the low-mass end for a flux-limited sample is sensitive to the {\em a priori} distribution of magnitude at a given stellar mass. We show in Section \ref{sec:mag} that this distribution is biased toward the faint end (top panels in Figure \ref{fig:mag}). One could underestimate the incompleteness if this bias is not properly accounted for. {Third, measurement uncertainties in stellar mass will introduce contamination from low-mass galaxies in a given mass bin, and thereby lead to an overestimate of their number density, especially at the high-mass end where the stellar mass function is steep \citep[e.g.][]{davidzon.2017:cosmos.2015.smf}.} In addition, cosmic variance may also lead to discrepancies at the massive end.

Nevertheless, our simulations do not include halos more massive than $\Mhalo\sim10^{12}\,\Msun$ and only include a small number of independent halos above $\Mhalo\sim10^{11}\,\Msun$. We may underestimate the scatter of certain galaxy properties at these masses. Moreover, we do not consider primordial chemistry or the ionizing background fluctuation prior to reionization, which may have important effects on halos below $\Mhalo\sim10^8\,\Msun$. Our predictions should be tested by future observations to better understand the uncertainties in the current model.

\begin{figure}
\centering
\includegraphics[width=\linewidth]{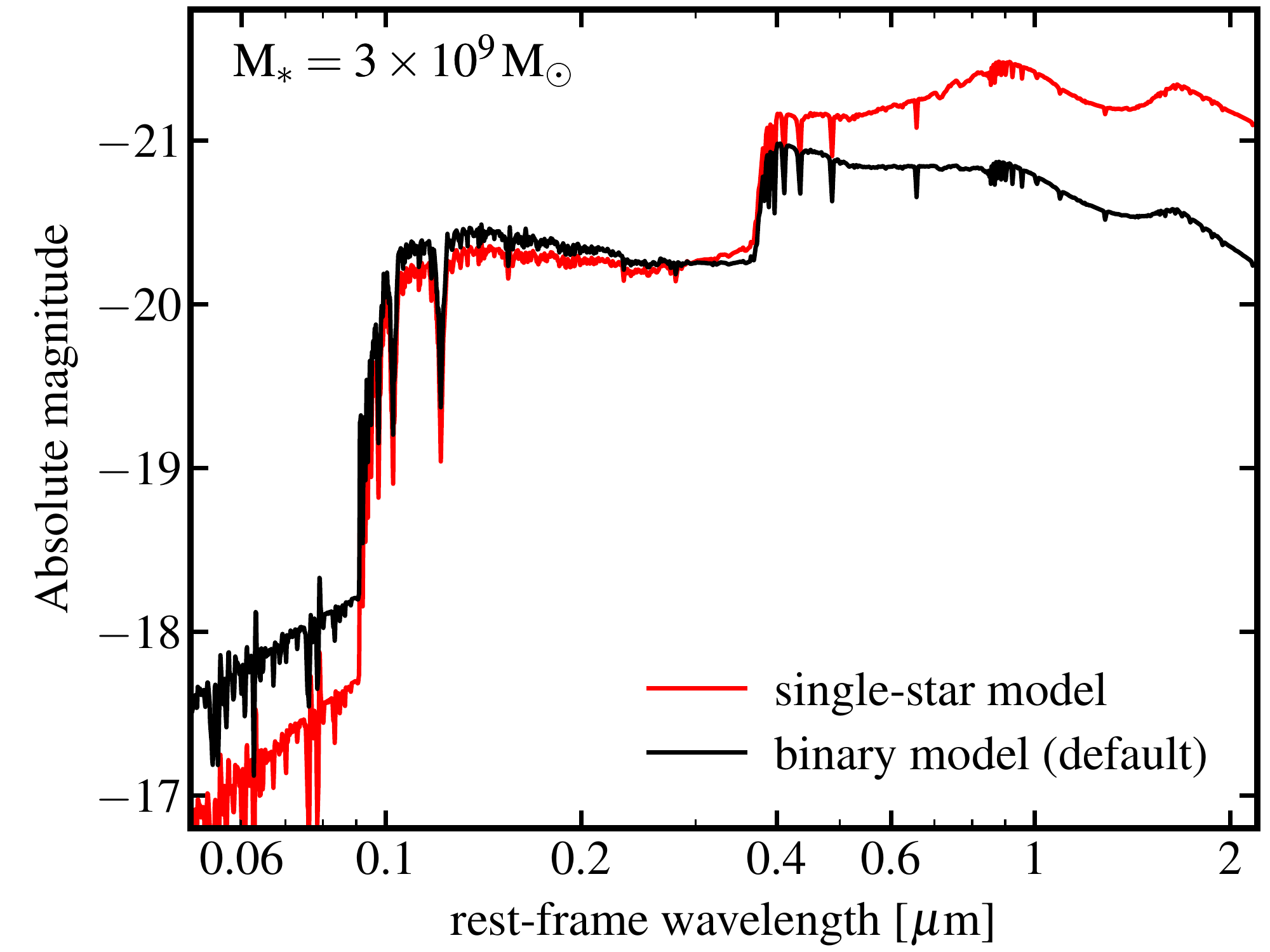} 
\caption{Synthetic spectrum for a $\Ms=3\times10^9\,\Msun$ galaxy (simulation z5m12a at $z=5$) using different stellar population models. Binary models produce slightly higher rest-frame UV luminosities at 1500\,\AA, but approximately 0.2--0.8\,mag lower rest-frame optical-to-IR luminosities than single-star models. Moreover, binary models produces more ionizing photons (wavelengths shorter than 912\,\AA), but these photons only contribute a small fraction of the bolometric luminosity.}
\label{fig:spectrum}
\end{figure} %

\subsection{Differences between stellar population models}
\label{sec:bpass}
In this paper, we use the BPASSv2.0 binary model with a \citet{kroupa.2002:imf} IMF from 0.1--$100\,\Msun$ as our default stellar population model for post-processing. To illustrate the difference between stellar population models, we show the synthetic spectrum for simulation z5m12a at $z=5$ (stellar mass $\Ms=3\times10^9\,\Msun$) in Figure \ref{fig:spectrum} using two models in BPASSv2.0: the default binary model (black) and the single-star model with the same IMF ({red}). Again, the spectra only include intrinsic star light without accounting for dust attenuation and line emission. The binary models produce slightly higher luminosities in the rest-frame UV at wavelengths bluer than the Balmer break at 3648\,\AA, but about 0.2--0.8\,mag weaker emission in the rest-frame optical and IR than single-star models. This is because of consequences of binary interaction: (a) the `effective' IMF is changed and (b) a large fraction of red supergiants are removed and replaced with hot stripped stars (J. J. Eldridge, private communication). These effects are particularly important in stellar populations younger than 1\,Gyr, which are dominant in galaxies at $z\geq5$ when the age of the Universe is comparable and even younger. If single-star models are used, the predicted B-band and J-band magnitudes will be brighter by 0.2\,dex and 0.5\,dex, respectively. Future observations of high-redshift galaxies at the rest-frame optical bands will provide more hints of the stellar populations in these galaxies. Another important difference is that binary models tend to produce more ionizing photons (wavelengths shorter than 912\,\AA). The production of these photons even extends to 30\,Myr after the formation of a stellar population (as opposed to 10\,Myr in single-star models), so these photons are more likely to escape the galaxy and play an important role in cosmic reionization \citep{ma.2016:fire.fesc.binary,gotberg.2017:binary.ionizing.spec}. However, ionizing photons only contribute a small fraction (less than 10\%) of the bolometric luminosity. Note that the differences between different stellar evolution calculations for non-rotating, single-star models (e.g. {\sc starburst99} and BPASS, using the same IMF and stellar atmosphere models) are much smaller than the effects of binaries.

\section{Conclusion}
\label{sec:conclusion}
In this paper, we present a suite of cosmological zoom-in simulations at $z\geq5$ covering the $z=5$ halo mass range $\Mhalo\sim10^8$--$10^{12}\,\Msun$. These are high-resolution simulations (100--$7000\,\Msun$ baryonic mass resolution) using physically motivated models of the multi-phase ISM, star formation, and stellar feedback from the FIRE-2 simulation suite of the FIRE project \citep{hopkins.2017:fire2.numerics}. These simulations provide useful guidance for future observations with JWST and next-generation ground-based telescopes. Our simulations are complementary to simulations of the first stars and low-mass galaxies at $z\gtrsim15$ with sophisticated primordial chemistry as well as large-volume simulations using empirically calibrated star formation and feedback models at much poorer resolution. 

By utilizing all properly resolved halos in each zoom-in region, we obtain a simulated sample containing hundreds of independent halos at a given redshift (Figure \ref{fig:nhalo}). We also include all snapshots (separated by about 20\,Myr in time) in our analysis to account for time variability in galaxy properties. At low halo masses (e.g. $\Mhalo\leq10^{11}\,\Msun$ at $z\sim6$ or $\Mhalo\leq10^{10.5}\,\Msun$ at $z\sim10$), our sample includes a large number of independent halos to account for halo-to-halo variance. At higher halo masses, our sample is small, so we may underestimate the scatter of galaxy properties due to halo-to-halo variance (cf. Sections \ref{sec:definition} and \ref{sec:smhm}).

We use the BPASSv2.0 binary stellar population models with a \citet{kroupa.2002:imf} IMF to compute the broad-band photometry from starlight in these galaxies. We also use analytic halo mass functions to assign each simulated galaxy a proper number density that reflects its relative abundance in the Universe. In this paper, we study the stellar mass--halo mass relation, star formation histories, the relation between broad-band magnitude and stellar mass, and stellar mass function and luminosity functions. Our main results include the following:

(i) The stellar mass--halo mass relation shows little evolution at redshift $z=5$--12 (Figure \ref{fig:smhm}). The best-fit median relation and $1\sigma$-scatter are $\log\Ms=1.58\,(\log\Mhalo-10)+7.10$ and $\Delta\log\Ms =\exp\,[-0.14\,(\log\Mhalo-10)-1.10]$ in the halo mass range $\Mhalo=10^{7.5}$--$10^{12}\,\Msun$ (Section \ref{sec:smhm}). The $\Ms$--$\Mhalo$ relation may bend at $\Mhalo>10^{12}\,\Msun$ (as is inferred at lower redshifts), but this regime is not probed by our simulations.

(ii) The relation between SFR, halo mass (stellar mass), and redshift can be best described by $\log\SFR=1.58\,(\log\Mhalo-10)+2.20\,\log\left(\frac{1+z}{6}\right)-1.58$ and $\log\SFR=1.03\,(\log\Ms-10)+2.01\,\log\left(\frac{1+z}{6}\right)+1.36$. The slopes of the SFR--$\Mhalo$ and SFR--$\Ms$ relations do not depend on redshift, but the average SFR at fixed halo mass (stellar mass) increases with increasing redshift by approximate 0.7\,dex from $z=5$ to $z=12$ (Figure \ref{fig:sfhz}).

(iii) The mean SFR for galaxies below $\Mhalo\sim10^8\,\Msun$ or $\Ms\sim10^4\,\Msun$ below $z\sim6$ drops significantly (Figure \ref{fig:sfhz}), because star formation is suppressed in low-mass galaxies by the ionizing background near the end of reionization (see also the left most panel in Figure \ref{fig:sfh}). About 50\% of the halos at $\Mhalo\sim10^8\,\Msun$ at $z\sim6$ are dark halos that contain no stars. Halos of similar masses above $z\sim7$ or halos more massive than $\Mhalo\sim10^{8.5}\,\Msun$ at any redshift continue normal star formation.

(iv) We provide the median and $1\sigma$-scatter for the magnitude--stellar mass relation and stellar mass--magnitude relation at rest-frame 1500\,\AA, B band, and J band (Table \ref{tbl:magfit} and Figure \ref{fig:mag}). Both relations have large scatter. We emphasize that the two relations are fundamentally different from each other. At fixed stellar mass, the distribution of magnitudes is set by the range of recent star formation histories. At fixed magnitude, the distribution of stellar mass is biased toward the low-mass end, due to the higher abundance of low-mass galaxies in the Universe (Section \ref{sec:mag}).

(v) We predict the stellar mass function and luminosity functions at rest-frame 1500\,\AA, B band, and J band from $z=5$--12 (Figures \ref{fig:smf} and \ref{fig:lf}). Our results are broadly consistent with current observations (Figures \ref{fig:comp}) and can be tested by future observations with JWST and next-generation ground-based telescopes. We make our predictions public for future use (see Appendix \ref{app:smf} for details).

(vi) Both the stellar mass function and luminosity functions show steepening low-mass-end or faint-end slopes with increasing redshift (from $\alpha=-1.85$ at $z\sim6$ to $\alpha=-2.18$ at $z\sim12$, Figures \ref{fig:smf} and \ref{fig:lf}), as inherited from the steepening of the low-mass-end slope of the halo mass function.

(vii) The stellar mass function slightly flattens below $\Ms\sim10^{4.5}\,\Msun$ at $z\sim6$. This results from the high dark halo fraction at $\Mhalo\sim10^8\,\Msun$, due to star formation being suppressed by the ionizing background at these redshifts. Similarly, the $z=6$ luminosity functions also show a flattening at magnitudes fainter than $\rm M_{1500}\sim-12$, $\rm M_B\sim-12$ and $\rm M_J\sim-12$ (Section \ref{sec:smf}). There is no such flattening at higher redshifts.

(viii) We derive the star formation rate and stellar mass density at $z=5$--12 (Figures \ref{fig:sfrd} and \ref{fig:smd}). Our results are in good agreement with current observational constraints at $z\leq8$. At higher redshifts, both are dominated by low-mass galaxies. Future JWST observations can put more robust constraints on the mass assembly histories at these redshifts by measuring galaxy number densities below $\rm M_{UV}\sim15$ or $\Ms\sim10^8\,\Msun$.

(ix) Dust attenuation in the rest-frame UV becomes important for galaxies with intrinsic 1500\,\AA-magnitude brighter than $\rm M_{1500}\sim-20$ (Figure \ref{fig:extinction}). In our analysis, the bright-end shape of the UV luminosity function is primarily set by dust attenuation (Figure \ref{fig:comp}). Approximately 37\% (54\%) of the UV luminosity from galaxies brighter than $\rm M_{1500}=-13$ ($\rm M_{1500}=-17$) is obscured by dust at $z\sim6$.

We note the caveat that our simulations do not include primordial chemistry and $\rm H_2$ formation and dissociation, nor try to model Pop III star formation. These are important in understanding the cooling in primordial gas and metal enrichment at very high redshifts ($z\geq15$), which may affect the star formation efficiency in halos below $\Mhalo\sim10^8\,\Msun$ \citep[e.g.][]{wise.2014:reion.esc.frac,chen.2014:highz.scaling.relation}. Furthermore, we do not model cosmic reionization self-consistently; instead, we only apply a spatially uniform, redshift-dependent ionizing background. This ignores the fact that reionization is highly inhomogeneous \citep[e.g.][]{barkana.loeb.2001:reion.review,furlanetto.2004:reion.hii.growth,iliev.2006:reion.geometry,mcquinn.2007:reion.hii.morphology} and that even after reionization the ionizing background has large spatial fluctuations \citep[e.g.][]{becker.2015:patchy.reionization,davies.2016:large.fluctuation.uvb,daloisio.2018:uvb.fluctuations}. Preliminary results indicate that increasing the ionizing background strength by a factor of 10--100 may lower the stellar mass in halos of $z=5$ mass $\Mhalo\sim10^9\,\Msun$ by a factor of 2. This may lead to larger scatter in the stellar mass--halo mass relation at the low-mass end that we do not capture in the current study. These questions are worth further exploration.

These simulations have many applications. In the future, we will use them to study the size evolution of high-redshift galaxies, dust attenuation and IR luminosity functions, nebular line emissions, the escape fraction of ionizing photons, [C {\sc ii}] and CO luminosity functions, metal-enriched absorbers in the circum-galactic medium, Lyman-$\alpha$ radiative transfer, globular cluster formation, and more. We will also expand the simulation suite to lower and higher masses and more extreme environments at these redshifts.

\section*{Acknowledgments}
We thank J. J. Eldridge, Steven Finkelstein, Renyue Cen, Frank van den Bosch, Priya Natarajan, Avi Loeb, and Rychard Bouwens for helpful discussions.
The simulations used in this paper were run on XSEDE computational resources (allocations TG-AST120025, TG-AST130039, TG-AST140023, and TG-AST140064). 
The analysis was performed on the Caltech compute cluster ``Zwicky'' (NSF MRI award \#PHY-0960291).
Support for PFH was provided by an Alfred P. Sloan Research Fellowship, NASA ATP Grant NNX14AH35G, and NSF Collaborative Research Grant \#1411920 and CAREER grant \#1455342.
Support for SGK was provided by NASA through Einstein Postdoctoral Fellowship grant number PF5-160136 awarded by the Chandra X-ray Center, which is operated by the Smithsonian Astrophysical Observatory for NASA under contract NAS8-03060.
CAFG was supported by NSF through grants AST-1412836 and AST-1517491, by NASA through grant NNX15AB22G, and by STScI through grant HST-AR-14562.001.
EQ was supported by NASA ATP grant 12-APT12-0183, a Simons Investigator award from the Simons Foundation, and the David and Lucile Packard Foundation.
MBK was also partially supported by NASA through HST theory grants (programs AR-12836, AR-13888, AR-13896, and AR-14282) awarded by the Space Telescope Science Institute (STScI), which is operated by the Association of Universities for Research in Astronomy (AURA), Inc., under NASA contract NAS5-26555.
RF is supported by the Swiss National Science Foundation (grant No. 157591).
DK was supported by NSF grant AST-1412153, funds from the University of California, San Diego, and a Cottrell Scholar Award from the Research Corporation for Science Advancement.

\bibliography{ms}

\appendix

\section{The weighting method}
\label{app:weight}
\begin{figure*}
\centering
\includegraphics[width=\linewidth]{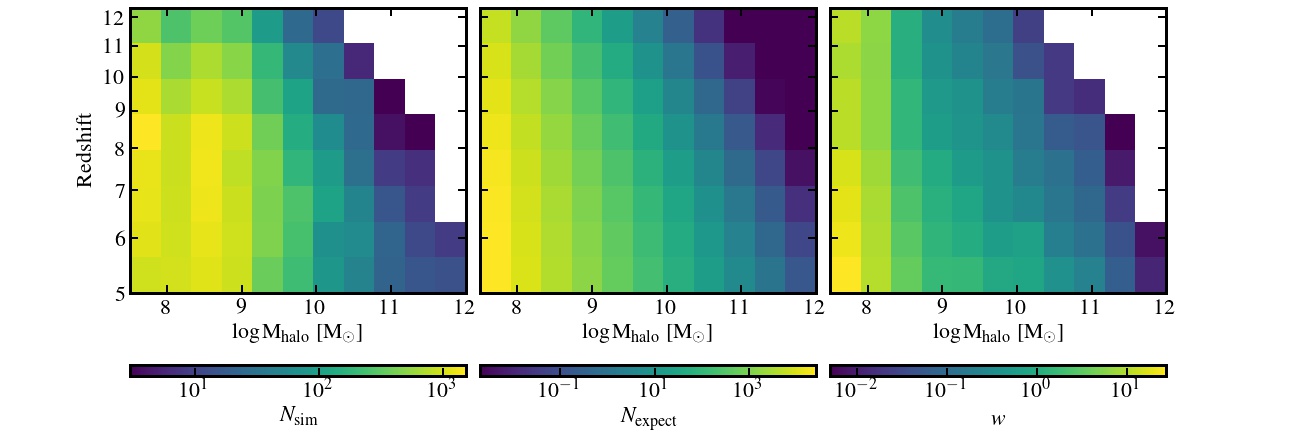}
\caption{The number of halo snapshots in our simulated catalog (left), the expected number of galaxies in the Universe (on $\rm 1\,arcmin^2$ sky, center), and the weight assigned to each halo snapshot (right) as a function of halo mass and redshift.}
\label{fig:weight}
\end{figure*}

In Section \ref{sec:weights}, we introduce a weighting method by assigning each halo `snapshot' a weight $w$ according to its halo mass and redshift to reflect its real abundance in the Universe. First, we bin our simulated catalog in $\log\Mhalo$--$\log(1+z)$ space with bin sizes $\Delta\log\Mhalo=0.4$ and $\Delta\log(1+z)=0.04$ and count the number of halo snapshots in each bin $N_{\rm sim}$ as shown in the left panel of Figure \ref{fig:weight}. Next, we compute the expected number of halos in the Universe in each bin $N_{\rm expect}=\phi\,\Delta\log\Mhalo\Delta V_{\rm com}$, where $\phi$ is the halo mass function obtained from {\tt HMFcalc} \citep[][number of halos per dex per comoving volume]{murray.2013:hmfcalc} and $\Delta V_{\rm com}$ is the comoving volume corresponding to the redshift range of each bin and $\rm 1\,arcmin^2$ area on the sky (this is to avoid $w$ being too large or too small). Each halo in the same bin will then be given the same weight $w=N_{\rm expect}/N_{\rm sim}$. Therefore, summing over the weights of halos in a given bin leads to the expected number of halos in the Universe (on $\rm 1\,arcmin^2$ area of the sky). We show $N_{\rm expect}$ and $w$ in the middle and right panel of Figure \ref{fig:weight}.

\section{Resolution tests}
\label{app:res}
\begin{figure}
\centering
\includegraphics[width=\linewidth]{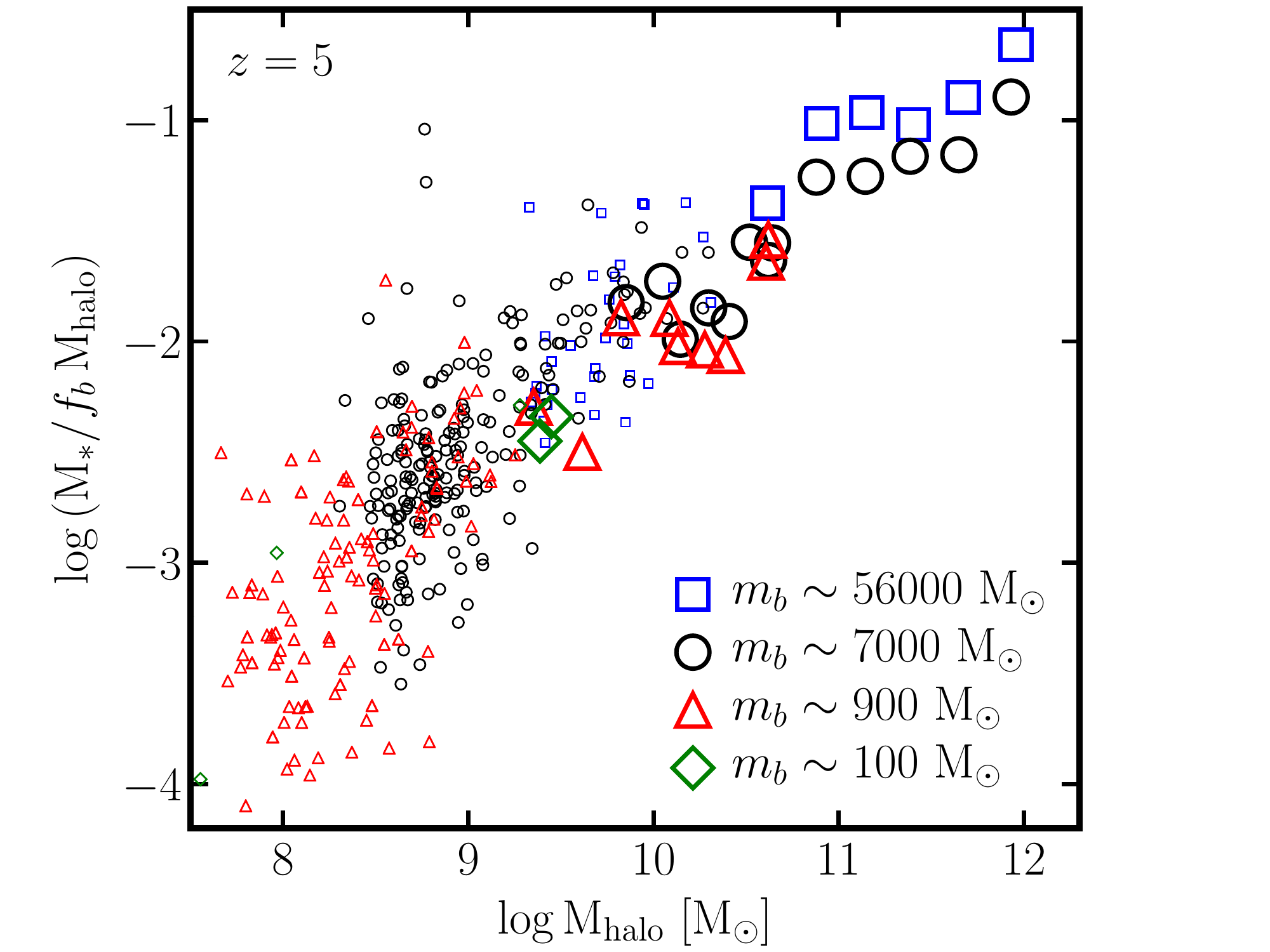}
\caption{The stellar mass--halo mass relation produced by simulations at different mass resolution. Simulations at mass resolution $m_b\sim5.6\times10^4\,\Msun$ systematically produce two times more stars. \referee{Simulations at resolution $m_b\sim7\times10^3\,\Msun$ and better do not show statistically-significant systematic differences over a fairly large number of galaxies (as we are able to show below $\Mhalo\sim10^{11}\,\Msun$). The difference in the stellar mass of individual galaxy is usually due to stochastic star formation and feedback.}} 
\label{fig:res}
\end{figure}

Every zoom-in simulation presented in the paper has been run at several resolution levels. The main text shows results only from the highest-resolution runs available. In Figure \ref{fig:res}, we show the stellar mass--halo mass relation in the $z=5$ snapshots for simulations using different mass resolution (shown by different symbols). The large symbols represent the most massive halo in each simulation and smaller symbols show less massive isolated halos in the zoom-in regions with more than $10^4$ particles and zero contamination. Note than we show the total stellar mass in the halo instead of the central galaxy stellar mass defined in Section \ref{sec:definition}, to reduce the effects of stochastic fluctuation in galaxy mass induced by mergers. Simulations at mass resolution $m_b\sim5.6\times10^4\,\Msun$ tend to systematically over-predict stellar mass by about a factor of two. This is also found in our previous work using ultra-high-resolution dwarf galaxy and Milky Way-mass galaxy simulations run with the same code down to $z=0$ \citep[see][]{wetzel.2016:high.res.mw.letter,hopkins.2017:fire2.numerics}. \referee{At resolution $m_b\sim7\times10^3\,\Msun$ and better, we do not find significant systematic differences in the stellar mass--halo mass relation for a fairly large sample of galaxies (as we show for $\Mhalo<10^{11}\,\Msun$). The difference in the stellar mass of individual galaxy is mainly due to stochastic effects: when and where a star particle forms and a SN occurs are stochastically sampled from the SFR and SNe rates. Any perturbations may affect the final stellar mass of each galaxy, but the statistics in the stellar mass--halo mass relation is unchanged. This is the way we define convergence for our simulations.} Therefore, we adopt mass resolution $m_b\sim7\times10^3\,\Msun$ for halos above $\Mhalo=10^{11}\,\Msun$ and even better resolution for our lower mass systems for final production runs, to ensure {reasonable} convergence and computational costs. For more extensive mass and spatial resolution tests, and other numerical details, see \citet{hopkins.2017:fire2.numerics}.

\section{Stellar mass functions and luminosity functions}
\label{app:smf}

In Sections \ref{sec:smf} and \ref{sec:lf}, we describe the method used to compute the stellar mass functions and luminosity functions from the simulated sample. Here in Tables \ref{tbl:smf} and \ref{tbl:lf}, we provide these results at $z=5$--12. The first two columns give the stellar mass functions above $\Ms=10^{3.5}\,\Msun$, and the remaining columns give the luminosity functions brighter than $\rm M_{AB}=-8$ at rest-frame 1500\,\AA, B, and J band, respectfully. In addition, we also make our model stellar mass function and luminosity functions public. A machine-readable version of these results is available at \href{http://www.tapir.caltech.edu/~xchma/data/hiz_smf_lf.zip}{http://www.tapir.caltech.edu/{\textasciitilde}xchma/data/hiz\_smf\_lf.zip}. Those derived from the simulated catalog are tabulated in files SMF\_sim\_zxx.txt,  LF\_UV\_sim\_zxx.txt, LF\_B\_sim\_zxx.txt, and LF\_J\_sim\_zxx.txt (these are identical to the data in Tables \ref{tbl:smf} and \ref{tbl:lf}). The model stellar mass functions and luminosity functions are tabulated in files SMF\_model\_zxx.txt,  LF\_UV\_model\_zxx.txt, LF\_B\_model\_zxx.txt, and LF\_J\_model\_zxx.txt (these are shown with the dashed lines in Figures \ref{fig:smf} and \ref{fig:lf}). The UV luminosity functions after accounting for dust attenuation are tabulated in LF\_UV\_red\_zxx.txt. The two digits xx in all file names represent the redshift. We encourage readers to use our results and confront them with future observations and other model predictions.

\begin{table*}
\caption{Stellar mass functions and luminosity functions at rest-frame 1500\,\AA, B, and J band from $z=5$--12.}
\begin{threeparttable}
\begin{tabular}{cccccccc}
\hline
\multicolumn{2}{c}{Stellar mass function} & \multicolumn{6}{c}{Luminosity function} \\
\hline
$\log\Ms$ & $\log\phi_{\ast}$ & M$_{1500}$ & $\log\phi_{1500}$ & 
M$_{\rm B}$ & $\log\phi_{\rm B}$ & M$_{\rm J}$ & $\log\phi_{\rm J}$ \\
($\Msun$) & (Mpc$^{-3}$\,dex$^{-1}$) & (mag) & (Mpc$^{-3}$\,mag$^{-1}$) & 
(mag) & (Mpc$^{-3}$\,mag$^{-1}$) & (mag) & (Mpc$^{-3}$\,mag$^{-1}$) \\
\hline
\multicolumn{8}{c}{$z=5$} \\
\hline
 3.83 & 0.76 & -22.37 & -3.96 & -22.18 & -3.95 & -22.60 & -4.05 \\
 4.50 & 0.62 & -20.68 & -3.01 & -20.52 & -2.94 & -20.89 & -3.04 \\
 5.17 & 0.33 & -18.99 & -2.42 & -18.85 & -2.37 & -19.17 & -2.36 \\
 5.84 & -0.04 & -17.30 & -1.84 & -17.18 & -1.73 & -17.45 & -1.76 \\
 6.51 & -0.47 & -15.61 & -1.14 & -15.51 & -1.05 & -15.73 & -1.08 \\
 7.18 & -0.92 & -13.92 & -0.69 & -13.84 & -0.67 & -14.01 & -0.66 \\
 7.85 & -1.62 & -12.23 & -0.38 & -12.17 & -0.22 & -12.30 & -0.20 \\
 8.52 & -2.35 & -10.54 & -0.09 & -10.50 & 0.11 & -10.58 & 0.15 \\
 9.19 & -2.55 & -8.85 & 0.20 & -8.83 & 0.42 & -8.86 & 0.47 \\
 9.85 & -4.02 &  &  &  &  &  &  \\
\hline
\multicolumn{8}{c}{$z=6$} \\
\hline
 3.83 & 0.87 & -22.34 & -4.10 & -22.01 & -4.18 & -22.34 & -4.19 \\
 4.48 & 0.62 & -20.66 & -3.35 & -20.37 & -3.35 & -20.66 & -3.35 \\
 5.13 & 0.35 & -18.97 & -2.64 & -18.72 & -2.51 & -18.97 & -2.64 \\
 5.78 & -0.09 & -17.28 & -1.84 & -17.07 & -1.81 & -17.28 & -1.86 \\
 6.43 & -0.56 & -15.59 & -1.25 & -15.42 & -1.27 & -15.59 & -1.29 \\
 7.08 & -1.12 & -13.91 & -0.80 & -13.77 & -0.70 & -13.91 & -0.70 \\
 7.74 & -1.66 & -12.22 & -0.41 & -12.12 & -0.30 & -12.22 & -0.27 \\
 8.39 & -2.32 & -10.53 & -0.06 & -10.47 & 0.11 & -10.53 & 0.13 \\
 9.04 & -2.88 & -8.84 & 0.24 & -8.83 & 0.42 & -8.84 & 0.48 \\
 9.69 & -3.98 &  &  &  &  &  &  \\
\hline
\multicolumn{8}{c}{$z=7$} \\
\hline
 3.81 & 0.97 & -21.39 & -3.89 & -21.07 & -4.14 & -21.40 & -4.34 \\
 4.42 & 0.60 & -19.82 & -3.17 & -19.54 & -3.05 & -19.82 & -3.23 \\
 5.04 & 0.32 & -18.24 & -2.54 & -18.00 & -2.45 & -18.24 & -2.47 \\
 5.65 & -0.16 & -16.67 & -1.92 & -16.46 & -1.88 & -16.67 & -1.92 \\
 6.26 & -0.63 & -15.09 & -1.26 & -14.92 & -1.25 & -15.09 & -1.32 \\
 6.88 & -1.20 & -13.52 & -0.75 & -13.38 & -0.74 & -13.52 & -0.75 \\
 7.49 & -1.68 & -11.94 & -0.37 & -11.85 & -0.26 & -11.94 & -0.25 \\
 8.11 & -2.32 & -10.36 & -0.04 & -10.31 & 0.05 & -10.37 & 0.13 \\
 8.72 & -2.99 & -8.79 & 0.26 & -8.77 & 0.46 & -8.79 & 0.50 \\
 9.33 & -4.00 &  &  &  &  &  &  \\

\hline
\multicolumn{8}{c}{$z=8$} \\
\hline
 3.86 & 0.87 & -21.46 & -4.31 & -20.90 & -4.22 & -20.91 & -4.22 \\
 4.57 & 0.48 & -19.39 & -3.21 & -18.92 & -2.95 & -18.93 & -2.96 \\
 5.29 & -0.01 & -17.32 & -2.25 & -16.93 & -2.27 & -16.94 & -2.23 \\
 6.00 & -0.61 & -15.25 & -1.43 & -14.95 & -1.45 & -14.95 & -1.45 \\
 6.72 & -1.31 & -13.18 & -0.72 & -12.96 & -0.70 & -12.97 & -0.68 \\
 7.43 & -1.89 & -11.11 & -0.17 & -10.98 & -0.14 & -10.98 & -0.11 \\
 8.15 & -2.64 & -9.04 & 0.32 & -8.99 & 0.45 & -8.99 & 0.47 \\
 8.86 & -3.89 &  &  &  &  &  &  \\
\hline
\multicolumn{8}{c}{$z=9$} \\
\hline
 3.83 & 0.77 & -20.85 & -4.95 & -20.27 & -4.93 & -20.33 & -4.93 \\
 4.50 & 0.39 & -18.87 & -3.14 & -18.39 & -3.16 & -18.43 & -3.17 \\
 5.17 & -0.11 & -16.90 & -2.38 & -16.50 & -2.25 & -16.53 & -2.27 \\
 5.84 & -0.69 & -14.92 & -1.52 & -14.61 & -1.51 & -14.64 & -1.56 \\
 6.50 & -1.38 & -12.94 & -0.80 & -12.72 & -0.83 & -12.74 & -0.84 \\
 7.17 & -2.00 & -10.97 & -0.20 & -10.83 & -0.22 & -10.85 & -0.22 \\
 7.84 & -2.69 & -8.99 & 0.20 & -8.94 & 0.36 & -8.95 & 0.36 \\
 8.51 & -3.66 &  &  &  &  &  &  \\
\hline
\multicolumn{8}{c}{$z=10$} \\
\hline
 3.80 & 0.68 & -19.25 & -3.73 & -18.50 & -3.60 & -18.44 & -3.58 \\
 4.41 & 0.32 & -17.52 & -2.90 & -16.88 & -2.79 & -16.84 & -2.70 \\
 5.01 & -0.12 & -15.79 & -2.18 & -15.27 & -2.15 & -15.23 & -2.23 \\
 5.62 & -0.74 & -14.06 & -1.41 & -13.65 & -1.40 & -13.62 & -1.40 \\
 6.22 & -1.39 & -12.33 & -0.75 & -12.04 & -0.78 & -12.02 & -0.78 \\
 6.83 & -1.98 & -10.60 & -0.21 & -10.42 & -0.23 & -10.41 & -0.18 \\
 7.43 & -2.54 & -8.87 & 0.16 & -8.81 & 0.29 & -8.81 & 0.27 \\
 8.04 & -3.44 &  &  &  &  &  &  \\
\hline
\end{tabular}
\begin{tablenotes}
\item Notes: the magnitudes are intrinsic magnitude without dust attenuation.
\end{tablenotes}
\end{threeparttable}
\label{tbl:smf}
\end{table*}

\begin{table*}
\caption{Table \ref{tbl:smf} --- continued.}
\begin{threeparttable}
\begin{tabular}{cccccccccc}
\hline
\multicolumn{2}{c}{Stellar mass function} & \multicolumn{6}{c}{Luminosity function} \\
\hline
$\log\Ms$ & $\log\phi_{\ast}$ & M$_{1500}$ & $\log\phi_{1500}$ & 
M$_{\rm B}$ & $\log\phi_{\rm B}$ & M$_{\rm J}$ & $\log\phi_{\rm J}$ \\
($\Msun$) & (Mpc$^{-3}$\,dex$^{-1}$) & (mag) & (Mpc$^{-3}$\,mag$^{-1}$) & 
(mag) & (Mpc$^{-3}$\,mag$^{-1}$) & (mag) & (Mpc$^{-3}$\,mag$^{-1}$) \\
\hline
\multicolumn{8}{c}{$z=11$} \\
\hline
 3.82 & 0.52 & -19.10 & -3.85 & -18.36 & -3.82 & -18.07 & -3.78 \\
 4.46 & 0.21 & -17.08 & -3.09 & -16.48 & -2.95 & -16.24 & -2.91 \\
 5.10 & -0.44 & -15.06 & -2.00 & -14.60 & -2.00 & -14.41 & -1.83 \\
 5.74 & -1.10 & -13.05 & -1.21 & -12.71 & -1.32 & -12.58 & -1.27 \\
 6.38 & -1.87 & -11.03 & -0.38 & -10.83 & -0.42 & -10.75 & -0.39 \\
 7.03 & -2.79 & -9.01 & 0.06 & -8.94 & 0.16 & -8.92 & 0.15 \\
 7.67 & -2.90 &  &  &  &  &  &  \\
\hline
\multicolumn{8}{c}{$z=12$} \\
\hline
 3.81 & 0.37 & -18.64 & -4.03 & -17.85 & -3.84 & -17.52 & -3.79 \\
 4.42 & 0.07 & -16.71 & -3.22 & -16.06 & -3.09 & -15.79 & -3.02 \\
 5.04 & -0.52 & -14.77 & -2.10 & -14.27 & -2.15 & -14.06 & -2.04 \\
 5.65 & -1.33 & -12.84 & -1.38 & -12.48 & -1.41 & -12.33 & -1.39 \\
 6.26 & -1.97 & -10.90 & -0.49 & -10.69 & -0.54 & -10.60 & -0.47 \\
 6.88 & -2.81 & -8.97 & -0.08 & -8.90 & 0.02 & -8.87 &  \\
 7.49 & -3.25 &  &  &  &  &  &  \\
\hline
\end{tabular}
\begin{tablenotes}
\item Notes: the magnitudes are intrinsic magnitude without dust attenuation.
\end{tablenotes}
\end{threeparttable}
\label{tbl:lf}
\end{table*}

\label{lastpage}

\end{document}